\RequirePackage{fix-cm}
\documentclass[smallextended]{svjour3}

\usepackage{multirow}
\usepackage{dblfloatfix}
\usepackage{bigstrut}
\usepackage[tikz]{bclogo}
\usepackage[skins]{tcolorbox}
\usepackage{wrapfig}
\usepackage[flushleft]{threeparttable}
\usepackage{float}
 
\definecolor{deeppink}{HTML}{D28986} 
\definecolor{pink}{HTML}{FFABA7} 
\definecolor{lightpink}{HTML}{FFCCC9} 
\usepackage[usestackEOL]{stackengine}
\usepackage{xcolor}

\definecolor{LightCyan}{rgb}{0.88,1,1}
\definecolor{darkgray}{gray}{0.7}
\definecolor{Gray}{rgb}{0.88,1,1}
\definecolor{Gray}{gray}{0.85}
\definecolor{Blue}{RGB}{0,29,193}
\definecolor{MyDarkBlue}{rgb}{0,0.08,0.45} 
\definecolor{pink}{rgb}{.96,.72,.77}
\definecolor{lightergray}{rgb}{0.85, 0.85, 0.85}
\definecolor{darkgray}{rgb}{0.47, 0.47, 0.47}
\definecolor{lightestgray}{rgb}{0.95, 0.95, 0.95}
\definecolor{ao(english)}{rgb}{0.0, 0.5, 0.0}
\definecolor{beige}{rgb}{0.96, 0.96, 0.86}
\usepackage{tcolorbox}
\newtcolorbox{blockquote}{colback=black!5,boxrule=0.4pt,colframe=black,fonttitle=\bfseries}
\newtcolorbox{RQbox}{colback=red!5,boxrule=0.4pt,colframe=black,top=1pt,bottom=1pt,fonttitle=\bfseries}

\usepackage{pifont}

\usepackage[final]{listings}
\usepackage{graphicx}
\usepackage{wrapfig}
\usepackage[hidelinks]{hyperref} 
\usepackage{alltt}
\usepackage{multirow}
\usepackage{graphicx}
\usepackage{color}
\usepackage[labelfont=bf,textfont={bf}]{caption}
\usepackage{subcaption}
\usepackage{pifont}
\usepackage{boxedminipage}
\usepackage{notoccite}
\usepackage{array}
\newcolumntype{C}[1]{>{\centering\let\newline\\\arraybackslash\hspace{0pt}}m{#1}}
\usepackage{enumerate}
\usepackage{booktabs}
\usepackage[framemethod=tikz]{mdframed}
\usepackage{lipsum}
\usetikzlibrary{shadows}
\usepackage{mathtools}
 \usepackage{paralist}
 \definecolor{light-gray}{gray}{0.80}
\usepackage{graphics}
\usepackage{longtable}
\newmdenv[
  tikzsetting= {fill=light-gray},
  linewidth=1pt,
  roundcorner=0pt, 
  shadow=false
]{myshadowbox}
\usepackage{colortbl} 
\usepackage{blindtext, graphicx}
\usepackage{textcomp}
\usepackage{mathtools}
\usepackage[final]{listings}
\usepackage{balance}
\usepackage{picture}
\usepackage{relsize}
\usepackage{multicol}
\usepackage{soul}
\usepackage{makecell}
\usepackage{bigstrut}

\definecolor{comment_color}{rgb}{0.5, 0, 1}
\definecolor{steel}{rgb}{0.1, 0.3, 0.5} 

\newcommand{\bi}{\begin{itemize}}
\newcommand{\ei}{\end{itemize}}
\usepackage{verbatim}
\usepackage{algorithmicx}
\usepackage{algpseudocode}
\usepackage[export]{adjustbox}
\definecolor{LightCyan}{rgb}{0.88,1,1}
\definecolor{darkgray}{gray}{0.7}
\definecolor{Gray}{rgb}{0.88,1,1}
\definecolor{Gray}{gray}{0.85}
\definecolor{Blue}{RGB}{0,29,193}
\definecolor{MyDarkBlue}{rgb}{0,0.08,0.45} 
\definecolor{pink}{rgb}{.96,.72,.77}
\definecolor{lightergray}{rgb}{0.85, 0.85, 0.85}
\definecolor{darkgray}{rgb}{0.47, 0.47, 0.47}
\definecolor{lightestgray}{rgb}{0.95, 0.95, 0.95}
\definecolor{ao(english)}{rgb}{0.0, 0.5, 0.0}
\definecolor{beige}{rgb}{0.96, 0.96, 0.86}
\usepackage{tcolorbox}

\usepackage{cite}

\usepackage{listings}
\definecolor{MyDarkBlue}{rgb}{0,0.08,0.45} 
\usepackage[linesnumbered,ruled,vlined]{algorithm2e}

\usepackage{cite}

\def\BibTeX{{\rm B\kern-.05em{\sc i\kern-.025em b}\kern-.08em
    T\kern-.1667em\lower.7ex\hbox{E}\kern-.125emX}}

\SetCommentSty{mycommfont}

\SetKwInput{KwInput}{Input}                
\SetKwInput{KwOutput}{Output}              

\SetKwFunction{FMain}{Main}
\SetKwFunction{FSum}{Sum}
\SetKwFunction{FSuba}{feature\_summarizer}
\SetKwFunction{FSubb}{cluster\_creator}
\SetKwFunction{FSubc}{bellwether\_finder}

\AtBeginDocument{%
  \providecommand\BibTeX{{%
    \normalfont B\kern-0.5em{\scshape i\kern-0.25em b}\kern-0.8em\TeX}}}

\newcommand{\IT}[1]{{\bf%
DODGE(\ifx*#1$\mathcal{E}$\else#1\fi)}}

\begin{document}
\title{When Less is More:   
On the Value of ``Co-training''
for Semi-Supervised   Software Defect Predictors}
\titlerunning{On the Value of ``Co-training''}

\author{Suvodeep Majumder, Joymallya Chakraborty, Tim Menzies}


\institute{S. Majumder \at
              Department of Computer Science, \\
              North Carolina State University,  Raleigh, USA \\
              \email{smajumd3@ncsu.edu} \\
              J. Chakraborty \at
              Department of Computer Science, \\
              North Carolina State University,  Raleigh, USA \\
              \email{chkrjoymallya@gmail.com}            
            \and 
            T. Menzies \at
              Department of Computer Science, \\
              North Carolina State University,  Raleigh, USA \\
              \email{timm@ieee.org}
}

\date{Received: date / Accepted: date}

\maketitle

\begin{abstract}
Labeling a module defective or non-defective is an expensive task. Hence, there are often limits on how much-labeled data is available for training. Semi-supervised classifiers use far fewer labels for training models. However, there are numerous semi-supervised methods, including self-labeling, co-training, maximal-margin, and graph-based methods, to name a few. Only a handful of these methods have been tested in SE  for (e.g.)   predicting defects-- and even there, those methods have been tested on just a handful of projects. 

This paper applies a wide range of  55 semi-supervised learners to over 714  projects. We find that semi-supervised ``co-training methods'' work significantly better than other approaches. Specifically, after labeling, just
 2.5\% of data, then make predictions that are competitive to those using 100\% of the data.
 
 That said, co-training needs to be used cautiously since the specific choice of co-training methods needs to be carefully selected based on a user's specific goals. Also, we warn that a commonly-used co-training method (``multi-view''-- where different learners get different sets of columns) does not improve predictions (while adding too much to the run time costs 11 hours vs. 1.8 hours).

 It is an open question, worthy of future work, to test if these reductions can be seen in other areas of software analytics. To assist with exploring other areas, all the codes used  are available at \url{https://github.com/ai-se/Semi-Supervised}.
\end{abstract}



\keywords{Semi-supervised Learning, SSL, Self-training, Co-training, Boosting methods, Semi-supervised preprocessing, Clustering-based semi-supervised preprocessing, Intrinsically semi-supervised methods, Graph-Based Methods, Co-forest, Effort Aware Tri-training}

\section{Introduction}
\label{sec:intro}

One crucial step in software analytics is finding the labels (or "ground truth") for training data, as supervised learning requires labeling all training data. Supervised learning algorithms can perform poorly if they do not have adequately labeled training data. Maintaining accurate and up-to-date project data is crucial for effective development and problem-solving. However, many software engineering projects suffer from the issue of poorly maintained data, where information about the project's structure, codebase, and documentation is poorly maintained for various reasons. For these reasons, we have begun to mistrust current labeling methods. Fueling that mistrust are comments like those from Tu et al.~\cite{tu2020better}, who  say that  current labeling methods are sometimes naive, often error-prone, and always expensive:

\bi
    \item {\em Naive:} Defect prediction researchers~\cite{catolino2017just, hindle2008large, kamei2012large, kim2008classifying, mockus2000identifying} often  label a commit as "bug-fixing" when the commit text uses words like "bug, fix, wrong, error, fail, problem, patch ." Vasilescu et al.~\cite{vasilescu2018personnel, vasilescu2015quality} warns that this can be somewhat ad-hoc, particularly if researchers just peek at a few results, then tinker with regular expressions to combine a few keywords.
    
    \item {\em Error-prone:}  When Yu et al.~\cite{9226105} explored labels from prior work exploring technical debt, they found that more than 90\% of labels marked as "false positive" was true. In the SE literature, there are similar reports where data labels error have corrupted the majority of the data for security bug labeling~\cite{ 9371393}; for labeling false alarms in static code analysis~\cite{10.1145/3510003.3510214}; or for software code quality~\cite{Shepperd13}.

    \item {\em Expensive:} In our work, we have studied approximately 714 software projects, including 476K commit files. After an extensive analysis, Tu et al.~\cite{tu2020better} proposed a cost model for labeling that data. Assuming two people checking per commit, that data would need three years of effort to label the data.
\ei

Building defect prediction models using incorrect labels results in sub-optimal models and would have poor performance. To address this issue, researchers have explored many different labeling methods. Here, we explore {\em semi-supervised learning} (SSL)~\cite{zhu2009introduction,zhu2005semi}  which is  a set of techniques that fall between unsupervised learning (that uses no labels~\cite{huang, zhong2004unsupervised, yang2016effort, zhang2016cross}) and supervised learning (that must label all training data~\cite{Ra13, commitguru, Kim08changes,catolino17_jitmobile,kamei2012large}).

There are many ways to implement SSL, such as GMM~\cite{goldberg2009multi, demiriz1999semi} that (a)~clusters all examples using the independent attributes, then (b)~labels one example per cluster, then (c)~copies that label to all other items in that cluster. We conjecture that software defect data is particularly suitable for this kind of reasoning. We say this since data from software projects often contains a much-repeated structure. When structures repeat, anything learned about one point should also apply to many nearby points. For more on this conjecture, see \S\ref{sslse}.

Nevertheless, because SE might be amenable to SSL, that still leaves many engineering issues about strictly how SSL should be applied. The GMM method (mentioned above) is but one of an extensive set of SSL methods. For example, here is a short sample of those methods (and for details on the following technical terms, see later in this paper);

\bi
    \item Li et al.~\cite{li2012sample} explored the coForest and AcoForest, a co-training method under wrapper-based methods.  
    
    \item Li et al.~\cite{li2020effort}  also used a co-training-based method named Effort-Aware  Tri-Training (EATT). 
    
    \item Fitting-the-confident-Fits (FTcF) is another semi-supervised algorithm used in defect prediction~\cite{lu2012software}, a self-training-based method. 
    
    \item ExtRF, as proposed by He et al.~\cite{he2016software}, is another example of a self-training-based semi-supervised method used for defect prediction. 
    
    \item Zhang et al.~\cite{zhang2017label} uses a label propagation algorithm, a graph-based semi-supervised algorithm, to find bugs in software systems. 
    
    \item Hysom, as proposed by Abaei et al.~\cite{abaei2015empirical}, is a semi-supervised model based on a self-organizing map and artificial neural network, which falls under unsupervised preprocessing.   
\ei

\begin{figure}[!t]
\centering
    \includegraphics[width=\linewidth]{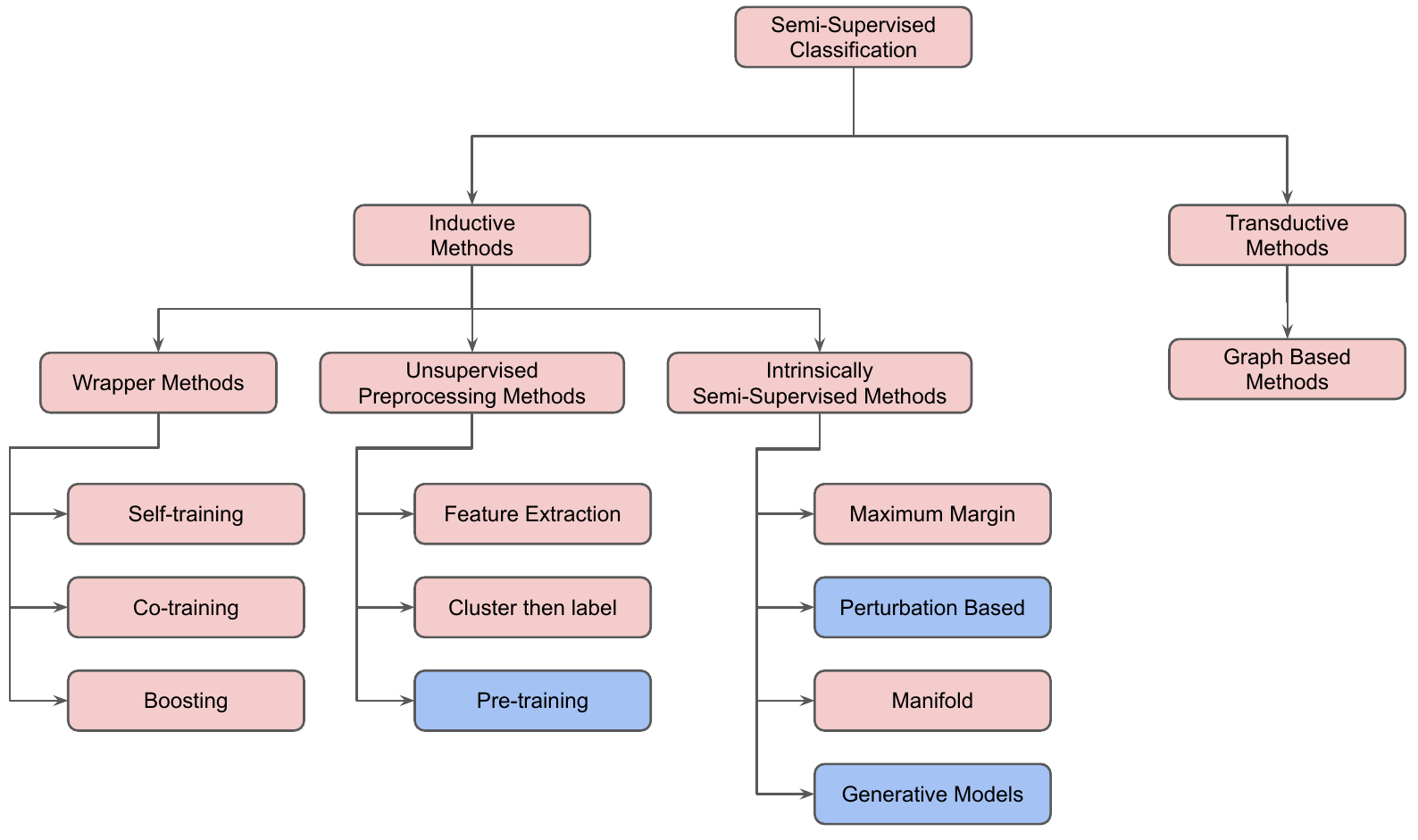}
    \caption{Taxonomy of Semi-Supervised learning from Van et al.~\cite{van2020survey}. This paper explores 55 methods from the pinks. The other blue nodes are left for future work since these notes use methods that are either  (a)~very computationally expensive or (b)~have been developed for data types not relevant to our target (defect prediction).} 
    \label{fig:semi_supervised_all}
\end{figure}

Figure~\ref{fig:semi_supervised_all} shows a taxonomy of Semi-Supervised learning from Van et al.~\cite{van2020survey}. The taxonomy shows that the SSL methods can be divided into  {\em inductive} and {\em transductive}, where the former attempts to find a classification model. At the same time, the latter tries to obtain label predictions for the given unlabelled data points. After that,  semi-supervised methods are divided based on how they utilize the unlabeled data points. This includes three categories,  {\em wrapper-based, up-supervised preprocessing, and intrinsically semi-supervised}. Inductive methods can be divided into Wrapper-Based, Unsupervised Preprocessing, and Intrinsically semi-supervised methods.

The starting point of this paper was the observation that even within all the above work:
\begin{quote}
    {\bf Most of the SSL work in SE comes from a small portion of the space of known algorithms from the left-hand side of   Figure~\ref{fig:semi_supervised_all}.}
\end{quote}

That is to say; there is much as-yet unexplored about SSL in SE. This paper studies 55 semi-supervised models on a   large dataset (714 projects) to answer the following research questions to mitigate that issue.

{\bf RQ1:} {\em Are there SSL models that perform better or worse than others?} We found that  an SSL method called {\bf co-training}   works best. In co-training, in repeated rounds,  one learner generates predictions using the most confident labels generated by the other.

{\bf RQ2:} {\em How do SSL models perform with increased data size?} This is the core question of this paper since unless we can show {\em good } performance with {\em limited} data, then this research is pointless. We found that using just  {\bf 2.5\%} of the labeled data is all we need for performance that is comparable (statistically similar) to supervised models that use 100\% of the labeled data. Pragmatically, this is the most important result of this paper since it means that SSL can reduce the effort of labeling by a factor of $100/2.5 = 40$.

Having established the main result of this paper, we then move on to cover some of the design choices within an SSL. The literature argues that different "views" of the  data~\cite{xu2013survey, iglesias2016hmm, hussain2016co} can generate better SSLs. For example, if we have feature weights that rank features 1,2,3,4,5,6, then Learner1 might only view features 1,3,5, while Learner2 might only view features 2,4,6. 

{\bf RQ3:} {\em Does the ``data view'' in co-training matter?}  At least for the data used here, we found that this advice is misguided since (a)~generating these data views can be computationally expensive; (b)~no performance gain was seen for any particular data view.

Another argument in the SSL literature\cite{xu2013survey, iglesias2016hmm, hussain2016co} is how to handle pseudo-labeling. The most straightforward policy is "self-teaching," where a learner {\em only} looks at the latest prior results from its previous iteration, while the other is "mutual-teaching", where pseudo-labeling involves learning from one or more different learners other than self.

{\bf RQ4:} {\em Does mutual-teaching improve SSL performance compared to self-teaching?}  We found that it is better to use pseudo labels generated from different learners (i.e., mutual teaching) than from a single learner (i.e., self-teaching).

From these results, we  summarize our contributions as follows: 

\begin{itemize}
    \item A experimental demonstration that specific SSL methods (single-view co-training)  can decrease the labeling cost of defect prediction by a factor of 100/2.5=40.
    \item A detailed empirical analysis and evaluation of semi-supervised algorithms and models to analyze these algorithms' capabilities and identify the best methods.
    \item One of the most extensive analyses yet seen in SE of semi-supervised algorithms (on 714     projects and 55 SSL methods). 
\end{itemize}

In order to allow other researchers to reuse/improve/refute our results, all our code and sample data are available online\footnote {\url{https://github.com/ai-se/Semi-Supervised}.}.
    
The rest of this paper is structured as follows. Section~\ref {sec:literature} discusses the background and related work. Our experimental methods are described in section~\ref{sec:Experimental_methods}. Data collection in section~\ref{sec:data} and learners used in this study are mentioned in section~\ref{sec:mylearners}, followed by the experimental framework in section~\ref{sec:experimental}. The evaluation criteria are in section~\ref{sec:eval}. The results and answers to the research questions are presented in section~\ref{sec:results}. Next, our result is discussed in section~\ref{sec:Discussion}. Threats to validity follow this in section~\ref{sec:threats}. Finally, a conclusion is in section~\ref{sec:conclusion}.

\section{Background   }
\label{sec:literature}

This section comments on the following:
\bi 
    \item Standard defect prediction methods (that need labels on all training data);
    \item Dozens of semi-supervised methods (that need far fewer labels);
    \item Our reasons why we believe SSL is particularly applicable to SE.
\ei
 
\subsection{Defect Prediction}
\label{sec:dp}

Bugs are not evenly distributed across   software~\cite{hamill2009common, koru2008investigation,  ostrand2004bugs, misirli2011ai}. Hence it is impractical and inefficient to distribute equal effort to every component in a software system~\cite{briand1993developing}. Algorithms that measure the criticality or bugginess of software using source code (product) or project history (process) are called defect prediction models. In a recent paper, Wan et al.~\cite{wan2018perceptions} reported much industrial interest in these predictors since the alternative is more time-consuming and expensive. Misirili et al.~\cite{misirli2011ai} and Kim et al.~\cite{kim2011empirical} report considerable cost savings when such predictors are used in guiding industrial quality assurance processes. Such defect predictors learned from static source code are remarkably effective compared to other methods. Rahman et al. ~\cite{rahman2014comparing} compared (a) static code analysis tools FindBugs, Jlint, and PMD with (b) defect predictors (which they called "statistical defect prediction") built using logistic regression. No significant differences in cost-effectiveness were observed. This is significant since defect prediction can be quickly adapted to new languages (by building lightweight parsers). At the same time, other methods (e.g., static code analyzers)   need extensive modification before they can be used in new languages. 

Due to their ease of use, defect predictors are widely applied:

\begin{enumerate}
    \item Application of defect predictors to locating security vulnerabilities~\cite{Shin2013}.
    \item Understanding what increases defects   (e.g., ratio comment to code, cyclomatic complexity) \cite{menzies10dp} or process metrics (e.g., see Table~\ref{tbl:process}).
    \item Predicting where the bugs are to allocate appropriate resources  (e.g.~\cite{bird09reliabity}).
    \item Using predictors to fix defects~\cite{arcuri2011practical} proactively.
    \item Release-level  change-level or just-in-time~\cite{commitguru} prediction.  
    \item  ``Transfer learning'' applies predictors built from one project to others~\cite{nam18tse}.
    \item Assessing different learning   predictors~\cite{ghotra15}. This has led to the development of hyperparameter optimization and better data harvesting tools.
\end{enumerate}

\begin{table}[!t]
\centering
\vspace{5mm}
\caption{List of process metrics used in this study}
\small
\begin{tabular}{r@{~:~}l}
adev     &  Active Dev Count                                 \\  
age      &  Interval between the last and the current change \\  
ddev     &  Distinct Dev Count                               \\  
sctr  &  Distribution of modified code across each file   \\
exp      &  Experience of the committer                      \\  
la       &  Lines of code added                              \\  
ld       &  Lines of code deleted                            \\  
lt       &  Lines of code in a file before the change        \\  
minor    &  Minor Contributor Count                          \\  
nadev    & Neighbor's Active Dev Count                      \\  
ncomm    & Neighbor's Commit Count                          \\  
nd       &  Number of Directories                            \\  
nddev    & Neighbor's Distinct Dev Count                    \\  
ns       &  Number of Subsystems                             \\ 
nuc      &  Number of unique changes to the modified files   \\  
own      & Owner's Contributed Lines                        \\  
sexp     &  Developer experience on a subsystem              \\  
rexp     & Recent developer experience \\
\end{tabular}
\label{tbl:process}
\end{table}

As to what data to collect, process metrics comment on "who" and "how" the code was written, while product metrics record "what" was written. Researchers and industry practitioners have tried many ways to identify essential features and why. Zimmermann et al.~\cite{zimmermann2007predicting} recommended complexity-based product metrics, and Zhou et al.~\cite{zhou2010ability} suggested size-based metrics, both of which are information regarding what was built. In contrast, developer-related metrics and change bursts metrics were recommended by Matsumoto et al.~\cite{matsumoto2010analysis} and Nagappan et al.~\cite{nagappan2010change}, respectively. Our metrics include the process metrics since prior studies have shown promising results and have been endorsed by Rahman et al.~\cite {rahman13}, Majumder et al.~\cite{majumder2022revisiting}, and others.

After saying "what" data to collect, the next question is, "how much" data to collect? In most of the above work, labels are needed for 100\% of the data. However, in many situations, labels are missing or mistrusted, meaning we must collect some new labels before the work can proceed. The lesson of this paper will be that, with the semi-supervised learning methods described below, we only need labels on a small part of the data (specifically, just 2.5\%).

\subsection{Semi-Supervised Learning}\label{ssl0}
Semi-supervised learning avoids searching labeled data or manually annotating unlabeled data. Such methods combine supervised and unsupervised learning, which uses a small amount of labeled data and a large amount of unlabeled data. 

According to Figure~\ref{fig:semi_supervised_all}, SSL methods can be divided into  {\em inductive} and {\em transductive}, where the former attempts to find a classification model. At the same time, the latter tries to obtain label predictions for the given unlabelled data points. After that,  semi-supervised methods are divided based on how they utilize the unlabeled data points. This includes three categories,  {\em wrapper-based, up-supervised preprocessing, and intrinsically semi-supervised}. Inductive methods can be divided into Wrapper-Based, Unsupervised Preprocessing, and Intrinsically semi-supervised methods. In comparison, transductive methods are graph-based methods.

\subsubsection{{Wrapper-Based Methods:}} Wrapper methods loop over the data and, in every loop, apply {\em training} followed by   {\em pseudo-labeling phase}. Starting with a very small of labels, the first training phase builds a model, which is then used to label the rest of the data. The labels that are "most confident" (defined below) are added to the pre-existing labeled data and used for training in the next loop. Our reading of the literature is that such wrappers are the most widely used algorithms in semi-supervised learning.

Wrapper-based methods can differ according to  how pseudo-labeling is performed:

\bi
    \item Self-training~\cite{scudder1965probability} methods use a supervised classifier that iterates and re-trains on its most confident predictions.
    \item In co-training~\cite{blum1998combining},  multiple supervised learners iterate and re-train on each other's most confident predictions.
    \item Boosting methods~\cite{zhou2012ensemble}  train ensemble methods sequentially by training on labeled data and the most confident predictions of the previous classifiers on unlabeled data. 
\ei

\subsubsection{{Semi-supervised preprocessing:}} Semi-supervised preprocessing uses two stages for building a semi-supervised model. In the first stage, the model uses hints from unlabeled data to extract or transform features or uses the features to form unsupervised clusters, while in the second phase uses the labeled data for training or labeling. Based on the first stage of semi-supervised preprocessing, they divided into two groups: 

\bi
    \item Feature extraction-based semi-supervised preprocessing~\cite{lu2012software}, where feature extraction is used to extract and transform the original data to a more linearly uncorrelated format. Such as principal component analysis(PCA)~\cite{wold1987principal} and Multidimensional scaling(MDS)~\cite{cox2008multidimensional}.
    \item Clustering-based semi-supervised preprocessing~\cite{goldberg2009multi, demiriz1999semi}, where labeled and unlabeled data are clustered and followed by uses of labeled data for using the resulting clusters to guide the classification process.
\ei

\subsubsection{{Intrinsically semi-supervised methods:}} In Intrinsically semi-supervised methods,  a decision boundary is formed using an objective function with two or more terms, using information from the labeled as well as information from the unlabelled examples found near the labeled data points. In this way, semi-supervised intrinsic learning can make do with a limited number of labels. These methods do not have base supervised learning steps as part of the process; instead, these are modified versions of different supervised algorithms with steps to consider unlabeled data. Most of these methods introduce some assumptions, such as low-density or smoothness, such as "maximum-margin''~\cite{vapnik1999overview, chapelle2005semi, li2014towards}. There are other ways to build intrinsically semi-supervised methods, including perturbation-based or Generative model-based methods; however, these methods mostly use computationally expensive deep neural networks and are excluded from further study in this paper.

\subsubsection{{Graph-Based Methods:}} The intuition with graph-based methods is that we do not need to label all data. Instead, given a few labels, the rest of the data can draw their labels from "nearby examples." Graph-based methods differ on (a)~how they find "nearby" examples and (b)~how they move those labels to themselves.

The methods generally involve three separate steps: graph creation, graph weighting, and inference~\cite{jebara2009graph, van2020survey}:
\bi
    \item First, data points are represented as a node in the graph connected based on some similarity measure. Consider the geometry of the dataset, which an empirical graph can represent $g = (V, E)$, where nodes $V = {1, ..., n}$ denote the training data and edges $E$ represent the similarities or affinity between adjacent nodes.
    \item Next, the resulting edges are weighted, yielding a weight matrix. 
    \item Once the graph is constructed, it is used for predicting the unlabeled data points. While iterating to generate the graph, it uses an objective function with two terms, one which penalizes predicted labels that do not match the actual label(for labeled examples), and another penalizes differences in the label predictions for connected nodes(for both labeled and unlabeled examples).
\ei

\subsection{SSL and SE}
\label{sslse}

Our introduction conjectured that data collected about source code is highly suitable for semi-supervised learning. This section presents some support for that conjecture. Note that if such repeated structure exists, if a learner labels any one thing, there should be many nearby similar things that deserve the same label. We hasten to add that the following is {\em suggestive}, but hardly {\em conclusive}, that software artifacts have repeated structures.   Nevertheless, in the following results, most examples could be labeled after examining just a few examples. For such data (as done in SSL), it should be possible to label a few examples and  propagate  those labels to the rest:
 
\bi 
    \item A repeated observation in the SE domain is that complex software can be controlled by characterized by just a few examples. For example, Yu et al. recently built a support vector machine model to find security vulnerabilities in 20,000+ Firefox functions~\cite{yu2019improving}. They found that around 200 support vectors were enough to define the boundary between vulnerable and non-vulnerable functions. 
    \item In other work, there are recent results~\cite{sucholutsky2021less} with "less than one'' -shot learning suggesting that it is possible and valuable to synthesize a small number of artificial exemplars by aggregating multiple examples. 
    \item Similarly, other examples demonstrate that generating tests only for the main branches in the code is an adequate testing strategy even for applications that process large cloud databases~\cite{zhang2020bigfuzz}. 
\ei
 
 As to why repeated structure can be seen in artifacts related to software projects, we believe it is because of {\em naturalness} and {\em power laws}:
 
\bi 
    \item {\em Naturalness:} Hindle and Devanbu wrote that "Programming languages, in theory, are complex, flexible and powerful, but the programs that real people write are mostly simple and rather repetitive, and thus they have statistical properties that can be captured in models and leveraged for SE tasks.~\cite{hindle2012naturalness}" To say that another way, computer programs are written using a programming language, and "language" is a technology that humans have been using for millennia to enable succinct communication. Repeated structures simplify communication since they let the observer learn the expected properties of a "typical" system. This means, in turn, they can recognize when some part of a system is anomalous  (because it lacks the usually repeated structures)~\cite{ray2016naturalness}. 
    \item {\em  Power-laws:} Code is written by people. The social interactions between people mean that those humans focus their work on tiny parts of the code (see Lin and Whitehead~\cite {Lin15}). To see this, consider a large system with modules A, B, C, and D. Developer 1 may only understand a small part of the code, e.g., module A. If Developer2 asks for help, then Developer1 will teach more about A than B, C, and D. When this cycle is repeated for Developer3, Developer4. Any activity log will contain similar, if not repeated, entries since it will record the activity of many people working on a small portion of the code. 
\ei

\section{Experimental Design}
\label{sec:Experimental_methods}
This section describes the methods used to evaluate 55 SSLs drawn from the above-mentioned methods.

\subsection{Data Collection}
\label{sec:data}
Until recently, many software analytics explored a small number of data sets passed from paper to paper(those data sets had names like XALAN or JEDIT). As a result, it was possible to question the external validity of conclusions reached in that manner since it only explored a small number of projects (often just five to 20).

These days, it is now possible to acquire a  much wider range of data via Github-based data collection. GitHub stores millions of projects; many are trivially very small, not maintained, or are not about -software development projects. To filter projects, we used the standard GitHub ``prudence checks'' recommended in the literature ~\cite{perils, curating}.

\begin{table}[!b]
\scriptsize
\begin{tabular}{rrrrrrr}
\rowcolor[HTML]{C0C0C0} 
\multicolumn{4}{c|}{\cellcolor[HTML]{C0C0C0}Process Metrics}                                                        & \multicolumn{3}{c}{\cellcolor[HTML]{C0C0C0}Data Statistics}                   \\
\rowcolor[HTML]{C0C0C0} 
\multicolumn{1}{c}{\cellcolor[HTML]{C0C0C0}Metric Name} & Median & \multicolumn{1}{c}{\cellcolor[HTML]{C0C0C0}IQR}  & \multicolumn{1}{c|}{\cellcolor[HTML]{C0C0C0}Range} & \multicolumn{1}{c}{\cellcolor[HTML]{C0C0C0}Data Property} & Median  & IQR     \\
la                                                      & 14     & \multicolumn{1}{r}{39}                           & \multicolumn{1}{r|}{[0,$+\inf$)}                   & Defect Ratio                                              & 37.60\% & 20.60\% \\
ld                                                      & 8      & \multicolumn{1}{r}{12}                           & \multicolumn{1}{r|}{[0,$+\inf$)}                          & Lines of Code                                             & 82K     & 200K    \\
lt                                                      & 92     & \multicolumn{1}{r}{122}                          & \multicolumn{1}{r|}{[0,$+\inf$)}                         & Number of Files                                           & 171     & 358     \\
age                                                     & 28.8   & \multicolumn{1}{r}{35}                           & \multicolumn{1}{r|}{[0,$+\inf$)}                          & Number of Developers                                      & 31      & 34      \\
ddev                                                    & 2      & \multicolumn{1}{r}{1}                            & \multicolumn{1}{r|}{[1,$+\inf$)}                           & Number of PRs.                                            & 55      & 101    \\
nuc                                                     & 6      & \multicolumn{1}{r}{3}                            & \multicolumn{1}{r|}{[0,$+\inf$)}                           & Number of Commits                                         & 217     & 379    \\
own                                                     & 1      & \multicolumn{1}{r}{0}                            & \multicolumn{1}{r|}{[0,$+\inf$)}                           & Duration                                                  & 186(W)  & 191(W) \\
minor                                                   & 0      & \multicolumn{1}{r}{0}                            & \multicolumn{1}{r|}{[0,$+\inf$)}                           & Number of Releases                                        & 20      & 32     \\
ndev                                                    & 27     & \multicolumn{1}{r}{22}                           & \multicolumn{1}{r|}{[0,$+\inf$)}                          & Number of Defective  Commits                              & 77      & 139     \\
ncomm                                                   & 71.1   & \multicolumn{1}{r}{49.5}                         & \multicolumn{1}{r|}{[0,$+\inf$)}                        & Number of Issues                                          & 46      & 67      \\
adev                                                    & 6      & \multicolumn{1}{r}{3}                            & \multicolumn{1}{r|}{[1,$+\inf$)}                           & Number of unique PR submitter                             & 5       & 6      \\
nadev                                                   & 72     & \multicolumn{1}{r}{49}                           & \multicolumn{1}{r|}{[0,$+\inf$)}                          &                                                           &         &         \\
avg\_nddev                                              & 2      & \multicolumn{1}{r}{2}                            & \multicolumn{1}{r|}{[0,$+\inf$)}                           &                                                           &         &         \\
avg\_nadev                                              & 7      & \multicolumn{1}{r}{5}                            & \multicolumn{1}{r|}{[0,$+\inf$)}                           &                                                           &         &         \\
avg\_ncomm                                              & 7      & \multicolumn{1}{r}{5}                            & \multicolumn{1}{r|}{[0,$+\inf$)}                           &                                                           &         &         \\
ns                                                      & 1      & \multicolumn{1}{r}{0}                            & \multicolumn{1}{r|}{[1,$+\inf$)}                           &                                                           &         &         \\
exp                                                     & 348.8  & \multicolumn{1}{r}{172.7}                        & \multicolumn{1}{r|}{[0,$+\inf$)}                       &                                                           &         &         \\
sexp                                                    & 145.7  & \multicolumn{1}{r}{70}                           & \multicolumn{1}{r|}{[0,$+\inf$)}                          &                                                           &         &         \\
rexp                                                    & 2.5    & \multicolumn{1}{r}{3.4}                          & \multicolumn{1}{r|}{[0,$+\inf$)}                         &                                                           &         &         \\
nd                                                      & 1      & \multicolumn{1}{r}{0}                            & \multicolumn{1}{r|}{[1,$+\inf$)}                           &                                                           &         &         \\
sctr                                                    & -0.2   & \multicolumn{1}{r}{0.1}                          & \multicolumn{1}{r|}{($-\inf$,$+\inf$)}                         &                                                           &         &        
\end{tabular}
\vspace{5mm}
\caption{Statistical median, IQR and range of values for the metrics used in this study (IQR denotes the (75-25)th percentile range) in Table~\ref{tbl:process}.}
\label{tbl:stats}
\end{table}

\begin{itemize}
    \item {\textit{{Collaboration}}: Refers to the number of pull requests, which must be more than one.}
    \item {\textit{{Commits}}: The project must contain more than 20 commits as recommended in the literature.}
    \item {\textit{{Duration}}: The project must contain software development activity of at least 50 weeks.}
    \item {\textit{{Issues}}: Project must contain more than ten issues.}
    \item {\textit{{Releases}}: Project must contain at least four releases.}
    \item {\textit{{Personal Purpose}}: The project must not be used and maintained by one person.}
    \item {\textit{{Software Development}}: The project must only be a placeholder for software development source code.}
    \item {\textit{{Defective Commits}}: Project must have at least ten defective commits.}
    \item {\textit{{Forked project}}:  Project must not be a forked project.}
\end{itemize}

The initial project pull was 5101 projects. After applying the prudence checks mentioned above,  we selected 714 projects\footnote{Some projects (27) had inconstant values for selected metrics and were removed from the selected project list}. The Data Statistics section of Table~\ref{tbl:stats} shows the median and IQR of each filtering criterion for the selected projects. We collected file-level process metrics for this research to answer our research questions. 

This data was extracted once and stored as pickle files in the following three steps:

\begin{enumerate}
    \item We collected 21 process metrics (following the definition either from commit\_guru or from the definitions shared by Rahman et al.) for each file in each commit by extracting the commit history of the project and then analyzing each commit for our metrics. We used a modified version of Commit\_Guru~\cite{rosen2015commit} code for this purpose, where instead of aggregating file-specific metric values for a commit, we store metric values for each file. We create objects for each new file we encounter and keep track of details (i.e., a developer who worked on the file, LOCs added, modified, and deleted by each developer) that we need to calculate. We also keep track of files modified together to calculate co-commit-based metrics. After collecting the 21 metrics mentioned in Table~\ref{tbl:stats} for each project, it is stored as a pickle file for prediction.
    \item Secondly, we use Commit\_Guru~\cite{rosen2015commit} code to identify buginducing and bugfixing commits. This process involves identifying bugfixing commits using a keyword\footnote{The keywords used are - \textit{bug, fix, error, issue, crash, problem, fail, defect, and patch}. These keywords are used by Rosen et al. in their commit\_guru~\cite{rosen2015commit} paper.} based search. Using these commits, the process uses the  commit\_guru's SZZ algorithm~\cite{williams2008szz, rosen2015commit} to find commits that were responsible for introducing those changes and marking them as buginducing \footnote{From this point onwards, we will denote the commit which has bugs in them as a ``bug-inducing''}. This process is performed on all commits throughout the project's life cycle. Note here that for a buginducing, each file labeled as a buggy file (bug-inducing) will have another instance of the same file, non-buggy (bugfixing). If a file has been fixed multiple times throughout the project history, it will have multiple instances in the dataset.
    \item  Thirdly, we used GitHub tag API to collect the release information for each project. We use the release number and release date information supplied from the API to group commits into releases and thus divide each project into multiple releases for each metric. Note here that we refer to a release number as the tags provided by the repository contributors, not by GitHub. Thus we apply regular expressions to match the release number to either ``XXXX'' or ``XXX'' format. A tag must differ in the section before the third dot to be considered a release.
\end{enumerate}


\subsection{Learners}
\label{sec:mylearners}
This section briefly explains the classification methods (supervised and semi-supervised) we used for this study. We selected the supervised methods following a prominent paper by Ghotra et al.'s~\cite{ghotra2015revisiting}. In defense of that selection, we note that  software engineering researchers widely use all these learners. For all the supervised models, we use the implementation from Scikit-Learn\footnote{https://scikit-learn.org/stable/index.html}. While for semi-supervised models, we have used the few available methods from Scikit-Learn, for the rest, we have either used the package created by the original authors of those methods or implemented the methods ourselves. 

\subsubsection{Supervised Learners:}
This section describes the supervised learners used in this study, either for the supervised learner or the base learner for wrapper-based methods. We have used the implementation and default parameters from Scikit-Learn. 

\begin{itemize}
    \item Support Vector Machine~\cite{ryu2016value,cao2018improved,tomar2015comparison}
    \item Naive Bayes~\cite{wang2013using,sun2012using,seiffert2014empirical,seliya2010predicting}
    \item Logistic Regression~\cite{ghotra2015revisiting,zhang2017data,he2012investigation,nam2013transfer,pan2010domain}
    \item Random Forest~\cite{tantithamthavorn2016automated, zhang2016cross, jacob2015improved, zhang2007predicting, ibrahim2017software, wang2013using, 9226105, h2022}
    \item K Nearest Neighbor~\cite{mabayoje2019parameter, goyal2022handling, iqbal2019performance, balogun2018software, gong2019empirical}
    \item Decision Tree~\cite{gayatri2010feature, pelayo2007applying, wang2012compressed, singh2017software, wahono2014comparison}
\end{itemize}

\subsubsection{Semi-supervised Learners:}
\label{sec:semi-supervised}
This section shows a detailed description of the semi-supervised methods used in this study. This includes a broader category of semi-supervised methods, which includes Wrapper-Based Methods, Semi-supervised preprocessing, Intrinsically semi-supervised methods, and Graph-Based Methods, as shown in figure~\ref{fig:semi_supervised_all}. All the following methods are available for download and execution from this paper's reproduction package \url{https://github.com/ai-se/Semi-Supervised}.

    \textbf{Self-training:} It is a type of semi-supervised learning method which falls under wrapper-based methods~\cite{yarowsky1995unsupervised, tanha2017semi}. This is a two-step process with multiple iterations. In the first step, the training phase, we train one of the supervised learners on the labeled data and possibly pseudo-labeled data from the last iteration. In the next phase, the pseudo-labeling phase, using the previous model, a portion of unlabeled data the learners are most confident of their predictions is labeled for use in the next iteration. 
    
    \textbf{Co-training:} This type of wrapper-based methods~\cite{goldman2000enhancing, zhou2004democratic, abney2002bootstrapping, balcan2004co} also contains two steps. In the training phase, two or more learners are trained on the labeled data and possibly pseudo-labeled data instead of one learner. The pseudo-labeled is based on the most confident prediction of the other learners~\cite{du2010does, goldman2000enhancing}. If there are more than two learners, the pseudo-labeling can be achieved by majority voting, booting, or other strategies~\cite{zhou2005tri, abney2002bootstrapping}. In Co-training, we adapt both single-view and multi-view settings. The key difference between the two models is that in the case of single-view models, both base learner sees the data with the same feature sets. During the pseudo-labeling phase for both base learners, the prediction labels and confidence scores are based on the same features. For multi-view co-training models, the two base learners are trained on different features, assuming that the features are not correlated. Thus, during the pseudo-labeling phase, when generating the prediction labels and confidence score, the models see completely different features (which should reduce over-fitting).
    
    \textbf{Effort Aware Tri-training (EATT)}: EATT is a type of co-training method that uses three learners. Zhou et al.~\cite{zhou2005tri} originally proposed the tri-training method as a traditional semi-supervised method. Zhang et al.~\cite{zhang2019effort} included effort-aware criteria for the pseudo-labeling phase in the tri-training method and achieved better performance both in terms of traditional metrics and effort-aware metrics. 
    
    \textbf{Co-forest:} It is an extension of the co-training algorithm, which incorporates Random Forest to tackle the problems of determining the most confident examples to label and produce the final hypothesis. Co-Forest first trains an ensemble of classifiers on the labeled data set, then during each iteration of Co-Forest, the training and pseudo labeling phase are performed~\cite{li2007improve, yu2010question}. In the case of the co-forest algorithm, during the pseudo-labeling phase, when determining the most confidently labeled examples for one of the classifiers of the ensemble, all other classifiers are used except the current one.

    \textbf{FTcF MDS:} It is an intrinsically semi-supervised method that combines a self-training algorithm with a dimension reduction technique. It was proposed by Lu et al.~\cite{lu2012software} in their study as a combination of semi-supervised preprocessing and self-training and is widely used~\cite{xu2016impact, ghotra2017large}. Here, Multidimensional Scaling(MDS)~\footnote{MDS calculates distances between each pair of points in the original high-dimensional space and then maps it to lower-dimensional space while preserving those distances between points as well as possible.} has been used as the preprocessing step. The model is built using the reduced dimension, where the error is minimized.

    \textbf{Semi-Boost:} It is a semi-supervised boosting algorithm~\cite{mallapragada2008semiboost, bennett2002exploiting}, which relies on optimization of an objective function consisting of two parts, one measuring the inconsistency between labeled and unlabeled examples and the other measuring the inconsistency among the unlabeled examples. The pseudo-labels are created using these two criteria. During training,  a new classifier is built, and a weight is assigned to the model using the labeled and pseudo-labeled data from previous steps.

    \textbf{S3VM:} Semi-Supervised Support Vector Machines (S3VM) is an widely used~\cite{liu2013adaptive, lee2007equilibrium} intrinsically semi-supervised method~\cite{vapnik1999overview, li2014towards}. In the case of traditional SVM, a supervised classification technique, it tries to find an optimal classification hyper-plane that meets the requirements of the classification task using an objective function. In the case of S3VM, another two terms are added with the objective function. One is a penalty term for the slack variable, which is the acceptable deviation between the function margin and the corresponding data. The other term is the loss function for unlabeled data. The S3VM is built by minimizing the objective, utilizing both labeled and unlabeled data simultaneously.

    \textbf{Semi-Supervised Clustering:} This is a semi-supervised preprocessing, specifically clustering then label method~\cite{goldberg2009multi, demiriz1999semi, bair2013semi}. Here, we use the Gaussian Mixture Model(GMM), where model parameters are estimated using the Expectation Maximization(EM) algorithm. In the first step, we use the GMM model provided by Scikit-Learn to build an unsupervised cluster using labeled and unlabeled data points. The tuning of the clusters is performed based on both two terms: BIC (Bayesian information criterion) and the error rate for the clusters. The final prediction for test data is performed using the cluster prediction followed by a label-assigning phase based on a cluster-to-label assignment.

    \textbf{Label Propagation:} This graph-based method creates a graph that connects instances in the training dataset and propagates labels through all data points. Each node changes its label to the one carried by the most significant number of its neighbors till convergence~\cite{zhu2002learning, xie2011community}.

    \textbf{Label Spreading:} Label spreading is also a graph-based algorithm proposed by Zhou et al.~\cite{zhou2003learning}. The algorithm works similarly to the Label propagation algorithm, except during the graph building, the similarity matrix is normalized before labels are assigned, and labels are changed in each iteration till the point the model converges by minimizing an objective function with two terms, one is the information from its neighbors, and the other is the initial information.

    \textbf{LSVM:} Laplacian Support Vector Machine(LSVM) is a semi-supervised support vector machine that uses manifold regularization and falls under intrinsically semi-supervised methods. Regularization is necessary to produce smooth decision functions and, thus, to avoid over-fitting the training data.

\subsection{Experimental Framework}
\label{sec:experimental}

\begin{figure}[]
\centering
    \includegraphics[width=\linewidth]{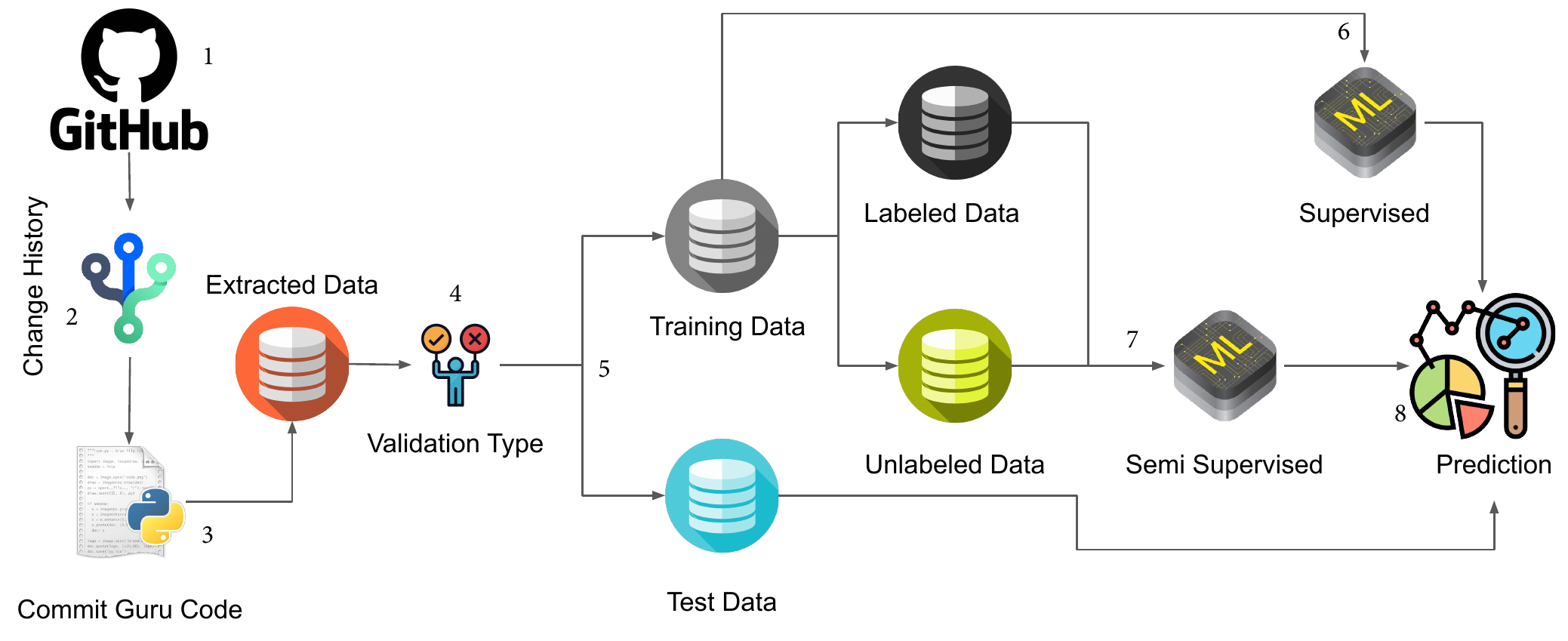}
    \caption{Framework} 
    \label{fig:framework}
\end{figure}

This section describes the experimental framework shown in Figure~\ref{fig:framework}. The framework consists of many sub-routines as follows -
\begin{enumerate}
    \item The process starts with selecting appropriate projects that meet the criteria by collecting meta-data about each project (collected using the GitHub search API) and using the GitHub filtering criteria mentioned in \S\ref{sec:data}.
    \item We use the filtered project list to collect 21 process metrics using the commit\_guru. We also collect release information using GitHub release API for each project using the process mentioned in \S\ref{sec:data}.
    \item We use the commit time for each file from the data collected using commit\_guru and release time from the release information to mark each row of the commit\_guru data with release numbers.
    \item We select which type of validation strategy to be used for the evaluation of the models.
    \item Based on the evaluation strategy, we divide the data into training and test data.
    \item Supervised models are trained with all available training data.
    \item We further divide the training data into labeled and unlabeled data for training semi-supervised models.
    \item The model performance is measured based on the test data.
\end{enumerate}

The evaluation of the models used in this study is based on two evaluation strategies described below - 

\begin{enumerate}
    \item {\em Cross-validation:} Here, we select all the files collected using the process described in Section~\ref{sec:data}. This includes the files labeled as buggy and non-buggy (this can include multiple copies of the same file if it was committed multiple times) throughout the project history. This data for each project is sorted randomly $M$ times. Then, the data is divided into $N$ stratified bins each time. Each bin, in turn, becomes the test set, and the remaining data is further divided into training and validation sets. For this study, we used $M=N=5$, repeated five times.
    \item {\em Release-based:} Here, we sort the data chronologically in releases $R_1, R_2,...$. Then we trained on data from release 1 to $R-3$, then tested on release $R-2$, $R-1$, and $R$. This temporal approach has the advantage that future data never appears in the training data.
\end{enumerate}

We use both strategies for completeness since both have been used widely in software analytics papers~\cite{tantithamthavorn2016empirical, koru2005building, yang2016effort, thota2020survey, bell2013limited, bennin2016empirical, hosseini2018benchmark}. As seen below (in Table~\ref{tbl:compare_r_c_1}), our conclusions are stable across both sampling strategies.

In order to handle imbalances in the class distribution, we  use the SMOTE algorithm to rebalance the training data for both  {\em cross-validation} and {\em release-based} after the labeled training data has been selected.

SMOTE (Synthetic Minority Oversampling Technique)~\cite{chawla2002smote} addresses data class imbalance issues. Imbalanced data is a problem since; if the target class is infrequent, it is hard for a learner to find the target. To mitigate class imbalance, SMOTE   samples from the minority classes and choosing $k$ nearest neighbors for each chosen sample. A synthetic instance is created at a randomly selected point between each pair of chosen samples and its neighbor. The synthetic samples are added to the original dataset to balance the ratio between majority and minority classes\footnote{  When performing such re-balancing, it is a methodological error to re-balance {\em both } the training {\em and} test sets (since learned models should be tested on data with the naturally occurring class frequencies. We assert that we re-balance {\em only} the training data and {\em not} the test.}.   We use SMOTE here since it has found prior success on our kinds of data~\cite{agarwal17}.

Before going on, we digress to make the following declaration.  Any learned model should be assessed on data distributions {\em that might appear naturally in the field}.  Hence, when using   SMOTE, it is a mistake to rebalance the {\em test} data.  Accordingly, we declare that while we rebalance the training data, {\em we leave the test data as is}.

\subsection{Evaluation Criteria}
\label{sec:eval}

This section describes the widely used ~\cite{kamei2012large, yang2016effort, yang2017tlel, xia2016collective, yang2015deep} evaluation metrics used in this study. These evaluation criterion has been designed to effectively measure the performance of machine learning models (specifically classification models). These include - 

\begin{itemize}
    \item Recall: Proportion of defective changes among all the defective changes
    \item Precision:  Proportion of defective changes among all the inspected changes
    \item False Alarm: Proportion of suggested defective changes that are not actual defective changes divided by everything that is not defective
    \item Popt20: Proportion of changes inspected by reading 20\% of the code
    \item IFA: Initial false alarms encountered before identifying the first defect.
    \item F1-Score: harmonic mean of the precision and recall.
    \item G-Score: harmonic mean of the (1 - false alarm) and recall.
\end{itemize}

Also, ideally, we want results that are stable across different experimental conditions. As mentioned above, we ran our rig twice: once as a cross-validation study and once again as a release-based study. We show below that our results are stable across those two conditions.

\subsection{Statistical Tests}
\label{sec:stats}

{\bf Scott-Knot Procedure:} Wee use the Scott-Knot test~\cite{mittas2013ranking, ghotra2015revisiting} to check if populations differ merely by random noise and an effect size test to check that two populations differ by more than a trivial amount. Scott-Knot is a  recursive bi-clustering method that executes over all the treatments, sorted by their median value.   If any two clusters are statistically indistinguishable, Scott-Knott assigns both the same ``rank''. These ranks have different interpretations, depending on whether we seek to minimize or maximize those numbers. For our purposes:

\bi
    \item Rank 1 is {\em worst} for recall, precision, Popt20 since we want to {\em maximize} these numbers.
    \item Rank 1 is {\em best} for the false alarm, and IFA since we want to {\em minimize} those.
\ei

\begin{wrapfigure}{r}{2in}
    \begin{center}
    \noindent\includegraphics[width=2in]{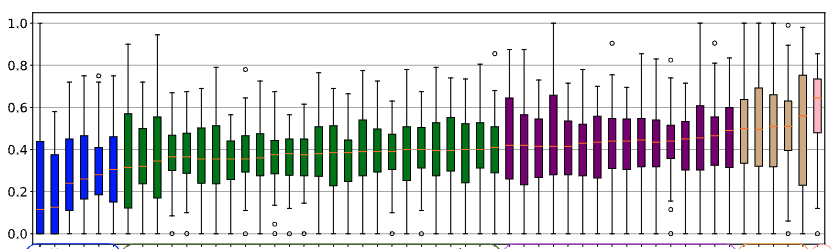} 
    Fig~\ref{fig:sk}a. The bunching effect (where many treatments have similar rankings).

    ~\\

    \includegraphics[width=2in]{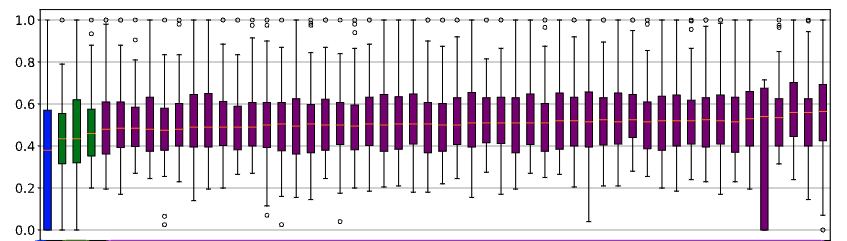} 
    Fig~\ref{fig:sk}b: The blurring effect (where many treatments have the same rank).
    \end{center}
    \caption{Examples of Scott-Knott results. In this figure, treatments with the same rank are assigned the same color.}\label{fig:sk}
\end{wrapfigure}

One benefit of using Scott-Knott is  {\em bunching}; i.e., a large number of treatments can be grouped into a small number of effectively similar units. For example, Figure~\ref{fig:sk}a shows one of our results (expanded below), where 55 different treatments became five similar groups. Bunching is useful since conclusions can be made with far less effort (over a few bunches) than if we tried to reason separately about (say) 55 treatments.   For example, in the sequel, we focus on treatments that do {\em not} exhibit {\em both} the worst false alarms  {\em and} recall. This reduces our space of treatments from the 55 seen in Figure~\ref{fig:sk}a  to the much more comprehensible set of nine treatments shown in Table~\ref{tbl:compare_r_c_1}.

Another effect seen with Scott-Knott is shown in Fig~\ref{fig:sk}b. When dealing with many treatments (as done here), the significant variances seen across all the treatments may mean that the results {\em blur}; i.e., that many of them are statistically indistinguishable. For example, in  Fig~\ref{fig:sk}b, we can see that 51 of the 55 treatments all receive the same rank. When such blurring occurs, all we can conclude is that the treatment explored is uninformative (in the sense that it does not distinguish the individual treatments).

\section{Results}
\label{sec:results}

\begin{figure}[]
\centering
    \includegraphics[width=\linewidth]{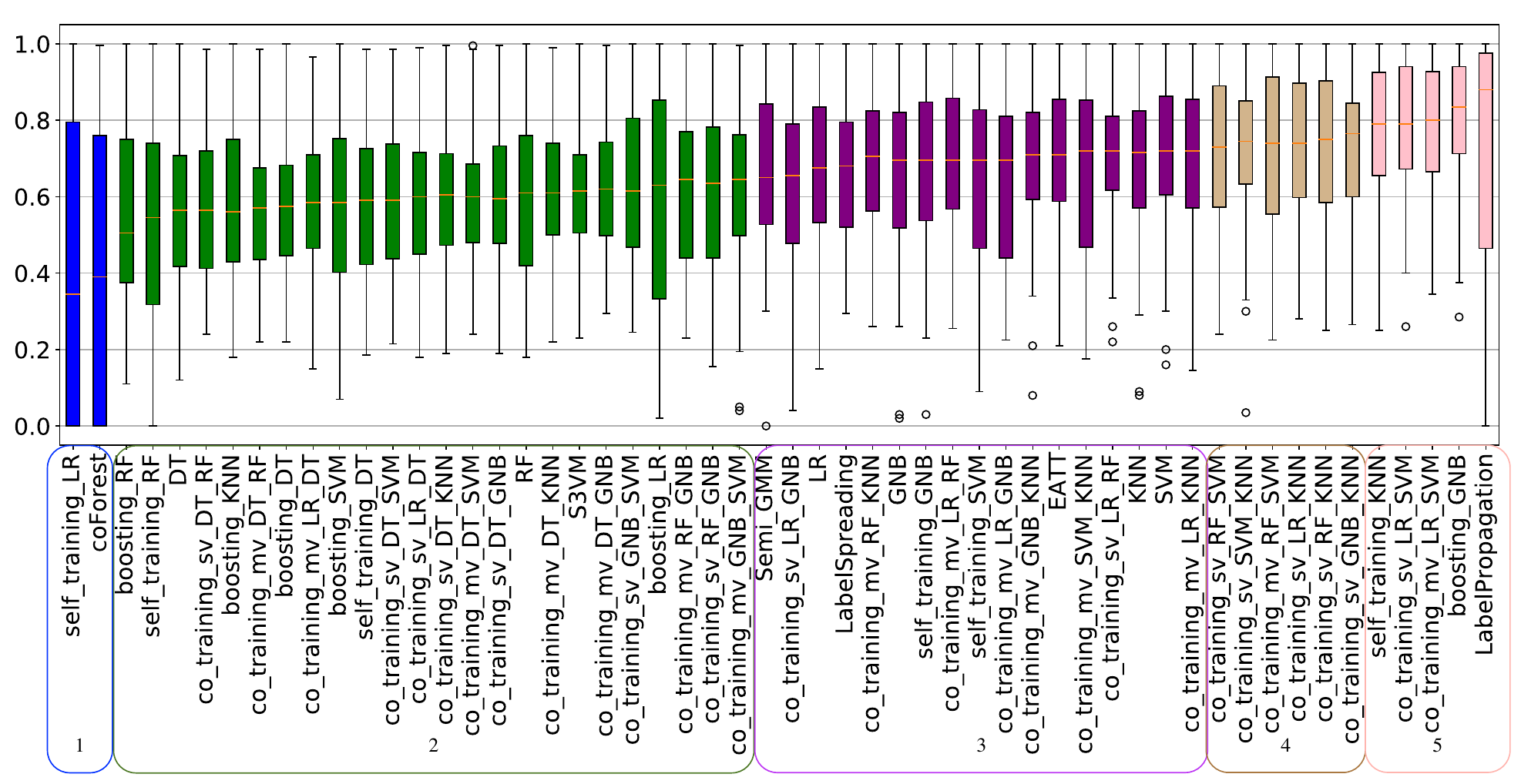}
    \caption{The Recall results are divided into five statistically different groups. Note that {\em larger} values are {\em better} for recall. The fully supervised methods (that use labels on 100\% of the data) are denoted DT, RF, LR, KNN, and SVM. Note that this group does not perform better or worse than the SSL methods (that use 2.5\% of the data). } 
    \label{fig:Recall}
    \vspace{0.3cm}
    \includegraphics[width=\linewidth]{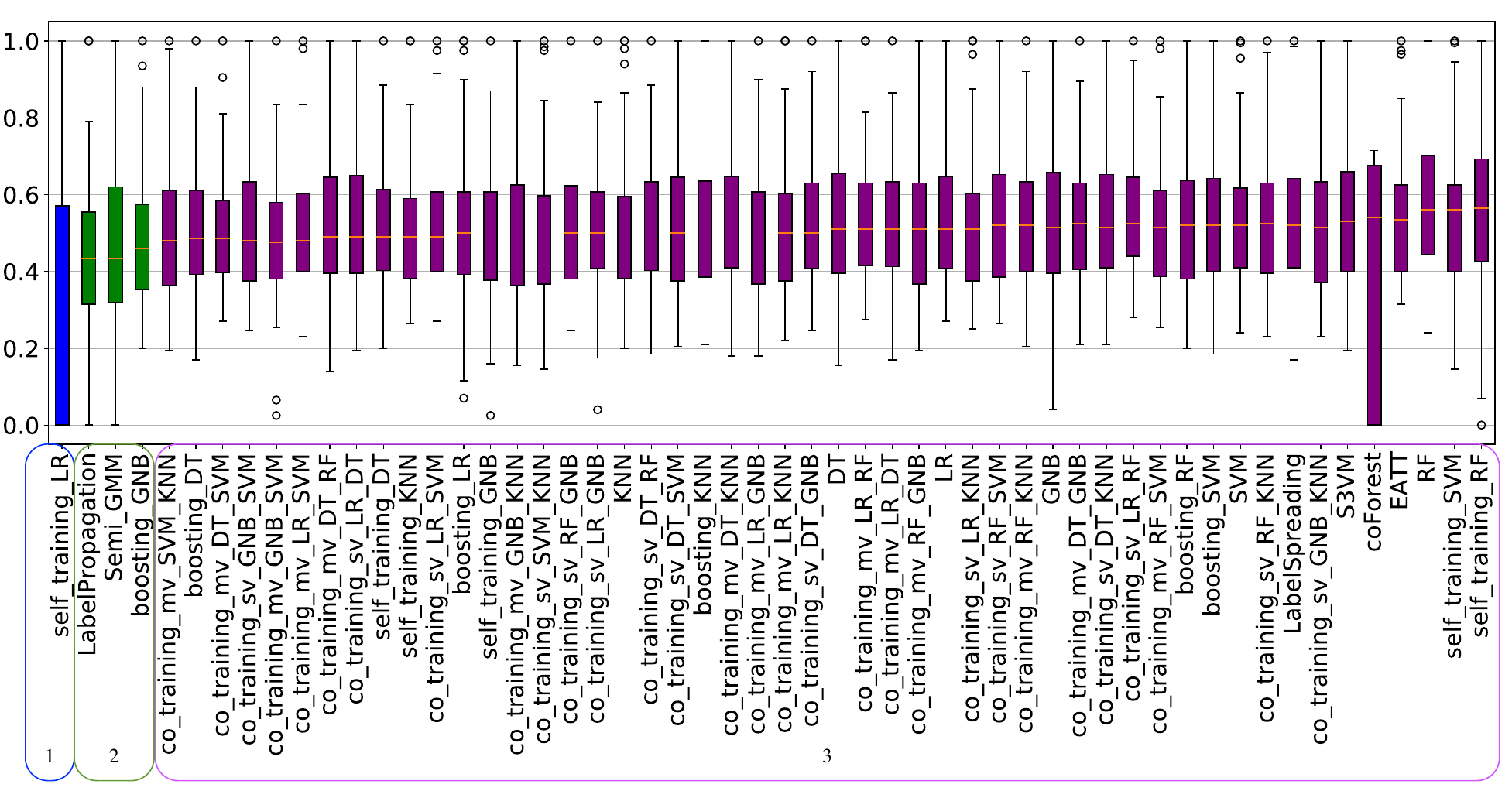}
    \caption{The precision results are divided into three statistically different groups. Note that {\em larger} values are {\em better} for precision. As with the recall results, the fully supervised methods  DT, RF, LR, KNN, and SVM do not perform outstandingly better or worse than the SSL methods.} 
    \label{fig:Precision}
\end{figure}

\begin{figure}[]
\centering
    \includegraphics[width=\linewidth]{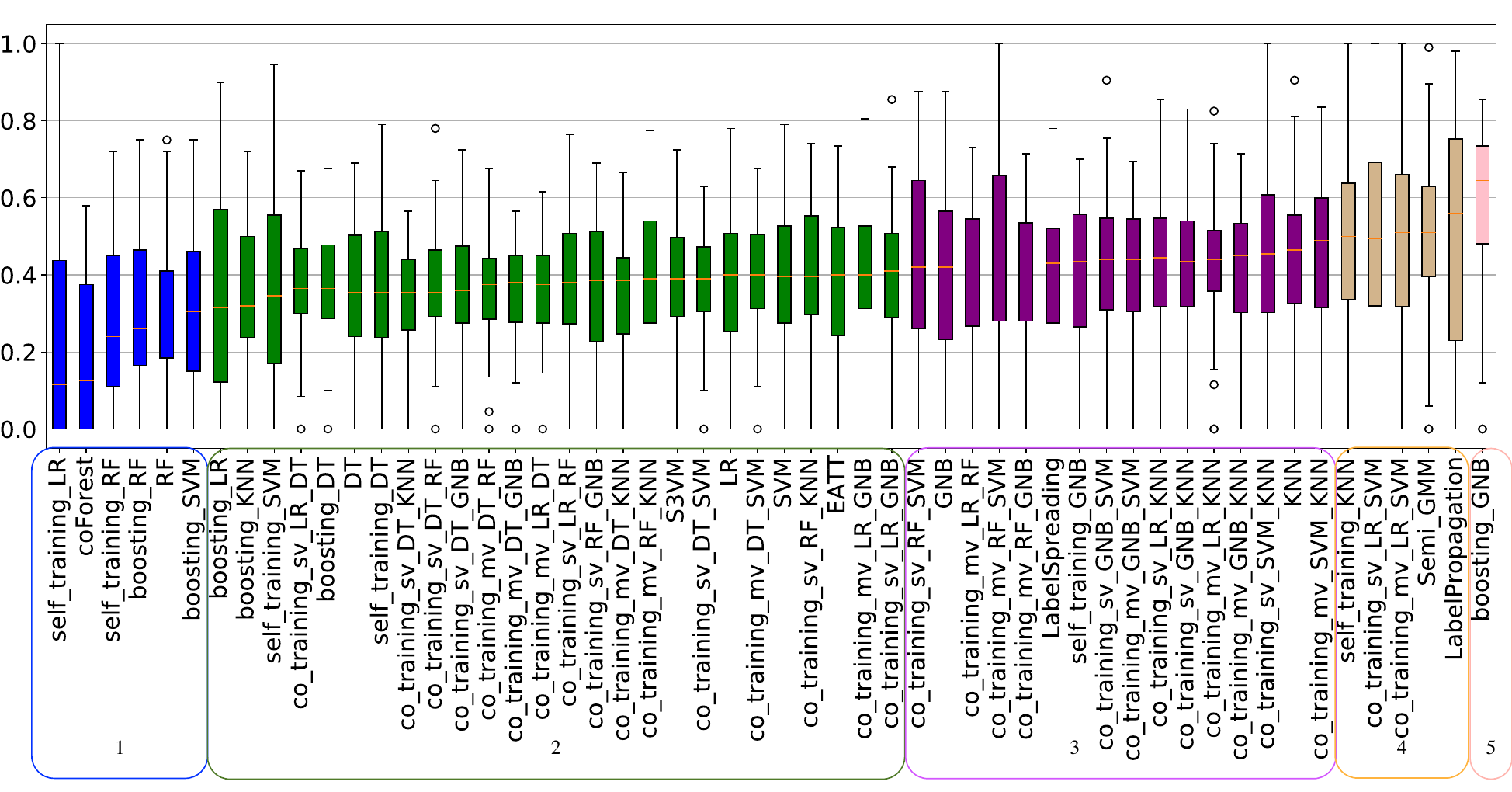}
    \caption{The false alarm results were divided into five statistically different groups. Note that {\em smaller} values are {\em better} for false alarm. Once again, we note that the fully supervised methods  DT, RF, LR, KNN, and SVM do not perform outstandingly better or worse than the SSL methods.
    } 
    \label{fig:Pf}
    \vspace{0.3cm}
    \includegraphics[width=\linewidth]{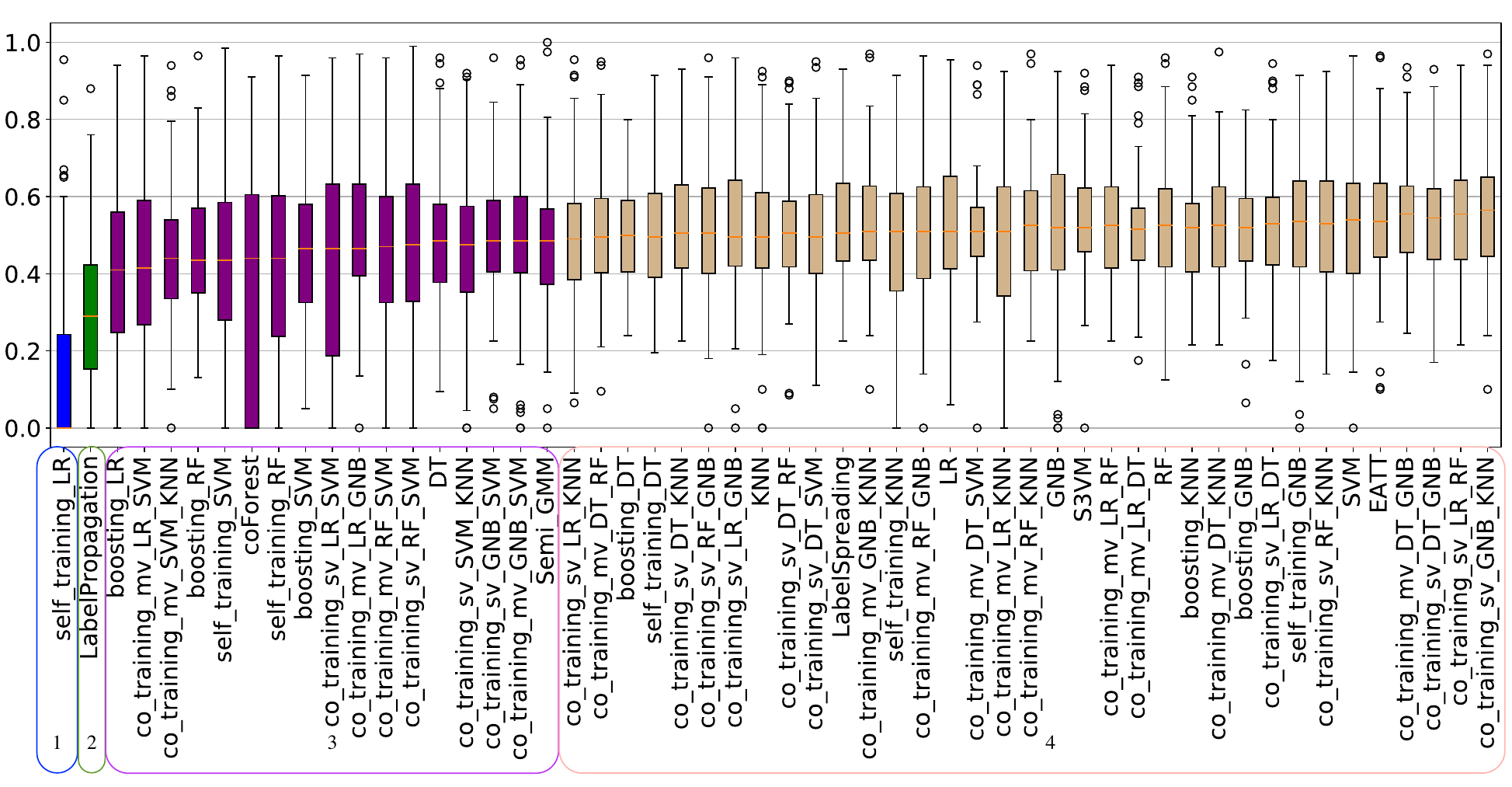}
    \caption{The g-score results are divided into four statistically different groups. Note that for g-score, {\em larger} values are {\em better}. As with all the other results, the fully supervised methods  DT, RF, LR, KNN, and SVM do not perform outstandingly differently from SSL methods.} 
    \label{fig:G-Score}
\end{figure}

This section describes our research questions and their findings in detail. We report recall, precision,   false alarm, and g-score in  Figure~\ref{fig:Recall} to  Figure~\ref{fig:G-Score}, respectively. Within each figure:

\bi 
    \item In those figures, supervised methods have short names (e.g., DT, RF, LR, KNN, and SVM), and unsupervised methods have much longer names (e.g., co\_training\_sv\_RF\_KNN)
    \item Each figure contains vertical box plots, where each box plot represents the performance of an individual model.
    \item Each box plot is generated using the respected performance evaluation criteria obtained using the evaluation strategy (this signifies that each box plot is generated with 25 data points). 
    \item  Each vertical plot is colored using the statistical methods of \S\ref{sec:stats}. Specifically, plots with the same color have the same rank (so plots with different colors are statistically distinguishable by more than a small effect).
\ei

In those results:
\bi  
    \item precision and g-score were {\em blurred}; i.e., nearly all these measurements for all these treatments were statistically indistinguishable. Hence we do not discuss those measures any further.
    \item Recall and false alarm were successfully {\em bunched} into 4 and 5 groups, respectively. After discarding the worst two groups in each set of results, we were left with nine treatments found in the top half of both recall and false alarm. Those treatments are shown in Table~\ref{tbl:compare_r_c_1}.
\ei 

As recommended by prior results of Tu et al.~\cite{tu2020better}, our first research question explores just 2.5\% of the labels. Subsequent research questions explored the effects of using more labels.

\subsection{\textbf{RQ1:Are there models which perform better or worse than others?}}
\label{sec:rq1}

Table~\ref{tbl:compare_r_c_1} shows the treatments selected by the above process. That table has two sets of results:   one for our cross-validation study and one more for our release-based study. In that  table, supervised methods are highlighted in gray. 

\begin{table}[!t]
    \centering
    \scriptsize
    \begin{tabular}{l|cc|cc|cc}
        \rowcolor[HTML]{9B9B9B} 
        \cellcolor[HTML]{9B9B9B}                                  & \multicolumn{2}{c}{\cellcolor[HTML]{9B9B9B}\textbf{Recall}} & \multicolumn{2}{c}{\cellcolor[HTML]{9B9B9B}\textbf{False Alarm}} \\
        \rowcolor[HTML]{9B9B9B} 
        \multirow{-2}{*}{\cellcolor[HTML]{9B9B9B}\textbf{Models}} & \textbf{Cross}                   & \textbf{Release}                   & \textbf{Cross}                      & \textbf{Rlease}  \\ \hline
        \rowcolor[HTML]{CDCAC9} LR                                & 0.69                         & 0.63                         & 0.4                             & 0.39  \\
        \rowcolor[HTML]{CDCAC9}SVM                                & 0.72                         & 0.7                          & 0.4                             & 0.38  \\\hline
        self\_training\_SVM                                       & 0.71                         & 0.65                         & 0.32                            & 0.34  \\\hline
        co\_training\_sv\_LR\_RF                                  & 0.72                         & 0.72                         & 0.38                            & 0.37  \\
        co\_training\_sv\_LR\_GNB                                 & 0.65                         & 0.66                         & 0.38                            & 0.38  \\
        co\_training\_sv\_RF\_KNN                                 & 0.75                         & 0.74                         & 0.37                            & 0.4   \\
        co\_training\_mv\_LR\_GNB                                 & 0.7                          & 0.67                         & 0.4                             & 0.4   \\
        co\_training\_mv\_RF\_KNN                                 & 0.74                         & 0.74                         & 0.39                            & 0.4   \\
        EATT                                                      & 0.71                         & 0.72                         & 0.4                             & 0.36  \\\hline
        \rowcolor[HTML]{FFCCC9}  Median & 0.71 & 0.7 & 0.39 & 0.38 \\
        \rowcolor[HTML]{FFCCC9}  Statistical Rank (Cross vs Release) & 1 & 1 & 1 & 1 
        \end{tabular}
        \caption{Comparison between Cross val and Release based evaluation for recall, false alarm, and g-Score. Cells with a gray background denote supervised methods; all other cells are semi-supervised. The models shown here are those that were ranked ``good'' in both  the recall results of  Figure~\ref{fig:Recall} and   the false alarm  results of Figure~\ref{fig:Pf} (aside:    precision and g-score results were not used due to the {\em blurring} explained in Figure\ref{fig:sk}b.). That is, this table only shows models that were  {\em not} blue or green in  Figure~\ref{fig:Recall}  and also which were {\em not} purple, tan or pink in  Figure~\ref{fig:Pf}. The last row of the table shows the statistical rank between cross-validation and release-based evaluation strategies.}
        \label{tbl:compare_r_c_1}
    \end{table}
    
\begin{figure}[!t]
     \centering
     \begin{subfigure}[b]{0.48\textwidth}
         \centering
         \includegraphics[width=\textwidth]{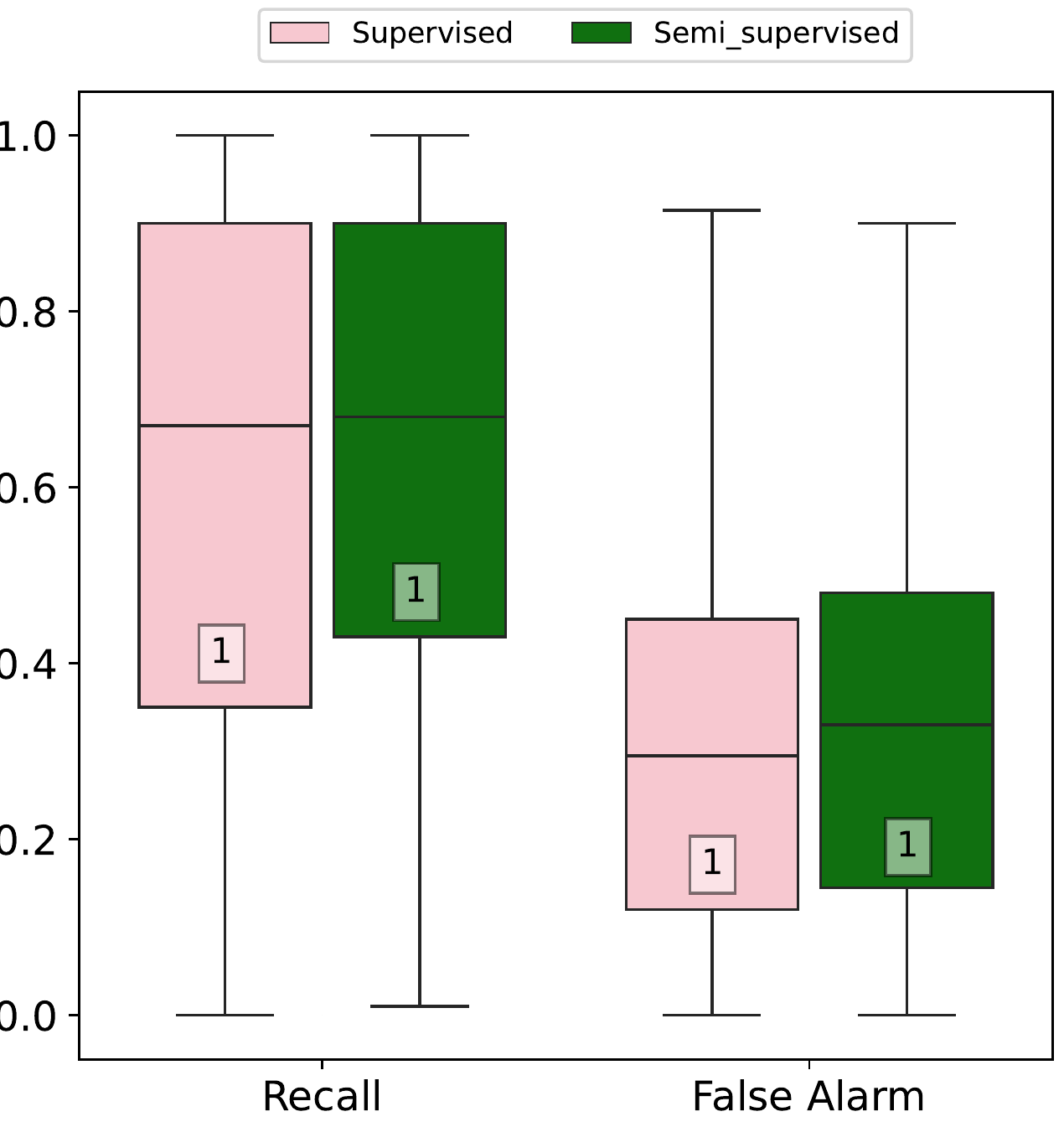}
         \caption{Supervised vs Semi\_supervised }
         \label{fig:supervsied_vs_semisupervised}
     \end{subfigure}
     \hfill
     \begin{subfigure}[b]{0.48\textwidth}
         \centering
         \includegraphics[width=\textwidth]{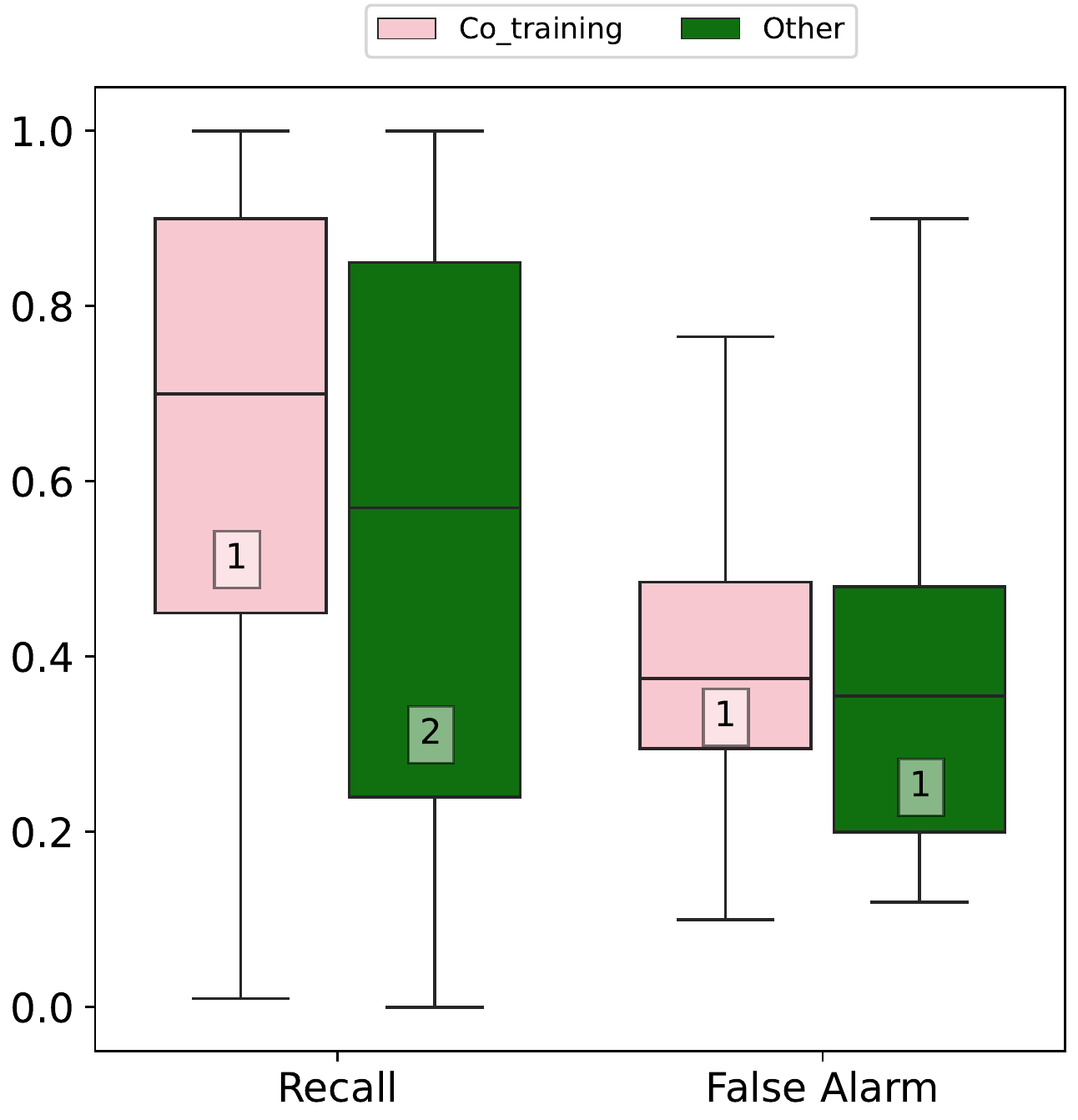}
         \caption{Co\_training vs Others}
         \label{fig:co_vs_other}
     \end{subfigure}
     \vspace{10mm}
     \caption{Comparing results between (a) supervised and semi\_supervised models and (b) co\_training and other non\_co\_training based models. The figure here shows the recall and false alarm of the two types. The number on the plots represents the ranking of the statistical tests for both performance measures between these two types of models.}
\end{figure}

We have subjected these results to our Scott-Knott statistical comparison:

\bi 
    \item  The bottom row of Table~\ref{tbl:compare_r_c_1} shows the ranks found in comparing the cross-based and release-based results. The ``1'' in that row indicates that the results from both sets of treatments were ranked the same; i.e., we can report that our results are {\em stable across both the cross-validation and release-based studies}.
    \item The numbers shown on the box plots of  Figure~\ref{fig:supervsied_vs_semisupervised} show the ranks found in comparing the supervised and the semi-supervised results. All those numbers are ``1'', meaning their performance is statistically indistinguishable (measured on false alarm and recall). 
\ei

For this paper, this second result (that SSL performs as well as full-supervised) is our most important finding since the semi-supervised results were achieved using just 2.5\% of the labeling effort required for supervised learning. This result makes us strongly recommend the use of SSL in SE.

As to what kind of semi-supervised model performs better overall,  Figure~\ref{fig:co_vs_other} applies Scott-Knott to compare co\_training with non\_co\_training based methods (denoted as ``Other'' in the table). We see that co\_training-based models perform statistically better than non\_co\_training-based methods. In contrast, they rank the same for false alarms. 

In summary, in answer to {\bf RQ1},
we say:

\begin{blockquote}
    In both cross-validation and release-based evaluation strategies, semi-supervised methods perform statistically the same, and thus, the choice of evaluation strategies does not significantly skew our analysis. Semi-supervised methods (that use only some data) can perform better than methods that use all the data. Finally, between different semi-supervised methods, co-training-based methods perform better than others.
\end{blockquote}

\subsection{\textbf{RQ2:How does models perform with increased data size?}}
\label{sec:rq2}

This section takes the SSL methods selected in Table~\ref{tbl:compare_r_c_1} and performs some extended analysis.

The previous research question concluded that all examples do not need labels on all the data. Given that result, the next question is, ``How much data can be ignored and still learn effective predictors?''. Note that, to reason about ``effectiveness,'' we must explore the benefits {\em and} the costs of labeling. Hence, our answer to {\bf RQ2} must discuss not just performance measures (like recall and false alarm rates) as the time required to achieve those results.

\begin{table}[!b]
    \centering
    \begin{tabular}{l|ccr}
    \rowcolor[HTML]{9B9B9B} 
    \textbf{percentage} & \textbf{\# files} & \textbf{time(hours)} & \textbf{Cost(\$)} \\ \hline
    2.5                 & 119,184            & 231                  & 3,811.5            \\
    5                   & 238,368            & 463                  & 7,639.5            \\
    10                  & 476,737            & 926                  & 15,279             \\
    20                  & 953,474            & 1,853                 & 30,574.5           \\
    100                 & 4,767,371           & 9,269                 & 152,938.5         
    \end{tabular}
    \caption{Time and Cost required for labeling effort. Based on a model
    from Tu et al.~\cite{tu2020better}.}
    \label{tbl:cost}
    \end{table}
    
    \begin{table}  
    \scriptsize
    \centering
    \begin{tabular}{l|cccc|cccc}
    \rowcolor[HTML]{9B9B9B} 
    \multicolumn{1}{c|}{\cellcolor[HTML]{9B9B9B}}                                 & \multicolumn{4}{c|}{\cellcolor[HTML]{9B9B9B}\textbf{Recall}}                                                & \multicolumn{4}{c}{\cellcolor[HTML]{9B9B9B}\textbf{False Alarm}}                                            \\
    \rowcolor[HTML]{9B9B9B} 
    \multicolumn{1}{c|}{\multirow{-2}{*}{\cellcolor[HTML]{9B9B9B}\textbf{Model}}} & \textbf{2.5\%} & \textbf{5\%}                 & \textbf{10\%}                & \textbf{20\%}                & \textbf{2.5\%} & \textbf{5\%}                 & \textbf{10\%}                & \textbf{20\%}                \\ \hline
    Self\_training\_SVM                                                           & 0.65           & 0.67                         & \cellcolor[HTML]{EFEFEF}0.75 & \cellcolor[HTML]{C0C0C0}0.76 & 0.34           & 0.3                          & \cellcolor[HTML]{EFEFEF}0.29 & \cellcolor[HTML]{EFEFEF}0.27 \\
    Co\_training\_sv\_LR\_RF                                                      & 0.72           & 0.72                         & \cellcolor[HTML]{EFEFEF}0.77 & \cellcolor[HTML]{EFEFEF}0.8  & 0.37           & \cellcolor[HTML]{EFEFEF}0.32 & \cellcolor[HTML]{EFEFEF}0.3  & \cellcolor[HTML]{C0C0C0}0.24 \\
    Co\_training\_sv\_LR\_GNB                                                     & 0.66           & \cellcolor[HTML]{EFEFEF}0.71 & \cellcolor[HTML]{C0C0C0}0.8  & \cellcolor[HTML]{9B9B9B}0.84 & 0.38           & 0.38                         & 0.38                         & 0.36                         \\
    Co\_training\_sv\_RF\_KNN                                                     & 0.74           & \cellcolor[HTML]{EFEFEF}0.76 & \cellcolor[HTML]{EFEFEF}0.76 & \cellcolor[HTML]{EFEFEF}0.8  & 0.4            & \cellcolor[HTML]{EFEFEF}0.35 & \cellcolor[HTML]{EFEFEF}0.32 & \cellcolor[HTML]{C0C0C0}0.28 \\
    Co\_training\_mv\_LR\_GNB                                                     & 0.67           & \cellcolor[HTML]{EFEFEF}0.73 & \cellcolor[HTML]{C0C0C0}0.8  & \cellcolor[HTML]{9B9B9B}0.84 & 0.4            & 0.4                          & 0.39                         & 0.38                         \\
    Co\_training\_mv\_RF\_KNN                                                     & 0.7            & 0.73                         & 0.74                         & \cellcolor[HTML]{EFEFEF}0.78 & 0.4            & \cellcolor[HTML]{EFEFEF}0.36 & \cellcolor[HTML]{C0C0C0}0.33 & \cellcolor[HTML]{C0C0C0}0.26 \\
    EATT                                                                          & 0.7            & 0.72                         & \cellcolor[HTML]{EFEFEF}0.8  & \cellcolor[HTML]{EFEFEF}0.82 & 0.36           & \cellcolor[HTML]{EFEFEF}0.32 & \cellcolor[HTML]{EFEFEF}0.29 & \cellcolor[HTML]{C0C0C0}0.27 \\ \hline
    Median                                                                        & 0.7            & 0.72                         & \cellcolor[HTML]{FFCCC9}0.77                         & \cellcolor[HTML]{FFCCC9}0.8                          & 0.38           & \cellcolor[HTML]{FFCCC9}0.35                         & \cellcolor[HTML]{FFCCC9}0.32                         & \cellcolor[HTML]{FFCCC9}0.27
    \end{tabular}
    \caption{Increased data size from 2.5\% to 20\% shown for the semi-supervised models selected in Table~\ref{tbl:compare_r_c_1} using the selection criteria mentioned. The cells highlighted in each row with different colors represent statistically different ranks.}
    \label{tbl:increased_data_size}
\end{table}

Table~\ref{tbl:cost}, we apply a   time/cost model to the labeling effort associated with learning from 2.5,5,10,20\% of the data. This model comes from Tu et al.~\cite{tu2020better} and assumes that labeling is conducted via crowd-sourcing (using Mechanical Turk); our crowd workers are being paid at least minimum age (\$8.25);  we assign two readers per issue report; and  50\% or our crowd workers produce poor results that need to be repeated (such a 50\% "cull rate" is common practice in crowdsourcing~\cite{chen2019replication}). To give these cost figures a little context (for university researchers), we note that many universities take a 50\% overhead on arriving research funds. Hence, paying for a 20\%  labeling   (\$30,574 at  50\% overhead equals    \$61,148)  would cost a university researcher about the same as hiring one more graduate research assistant for one year.

As to performance,  Table~\ref{tbl:increased_data_size} shows the impact on false alarm and recall of using   2.5\%, 5\%, 10\%, and 20\%  labeled data. As aside, we also note that we ran 1\% and 40\% and  did not include them  in Table~\ref{tbl:increased_data_size} since:

\bi  
    \item Running 1\% leads to results very much worse than 2.5\%
    \item Running 40\% leads to results statistically indistinguishable from 20\%.
\ei

In Table~\ref{tbl:increased_data_size}, cells are colored using a Scott-Knott analysis  (applied to the four treatments on each row). In that color scheme, white is {\em worst} and {\em darker} is {\em better}. From Table~\ref{tbl:cost} and Table~\ref{tbl:increased_data_size}, we make the following observations:

\bi 
    \item From {\bf RQ1}, we learned that SSL working on 2.5\% of the data works just as well as running on 100\% of the data. Table~\ref{tbl:increased_data_size} shows us that in between 2.5 and 100, there is a "sweet-spot" of 20\% labels which out-performs 2.5\% SSL (and hence, transitively, 100\% fully-supervised learning). 
    \item   On the other hand, from Table~\ref{tbl:cost}, we see that achieving that improvement from 20\% labeling increases the cost by a factor of $30,574.5/3,811.5\approx 8$. 
    \item From the last row of Table~\ref{tbl:increased_data_size}, we see that the size of the improvement from 2.5\% to 20\% is about ten points improvement in recall (median=0.7 to median=0.8) and 11 points in false alarm (median=0.38 to median=0.27).
    \item  Between 2.5\% to 20\%, there is a 10\% labeling policy that is four times as expensive (15279/3811) while offering some improvements over our 2.5 policy: about seven points improvement in recall (median=0.7 to median=0.77) and six points in false alarm (median=0.38 to median=0.32).
\ei

It is up to the reader to decide if a four to eight-fold increase in cost is worth  (e.g.)  seven to ten points more recall.   However they decide, our case that SSL is useful for SE will persist. If they decide that, on balance, that 2.5\% labeling is enough, then we can offer them a 40-fold reduction in the labeling effort. On the other hand, if they are willing to increase their labeling cost by a factor of four to eight, we can still offer them a 100/10=ten-fold or 100/20=five-fold reduction on the labeling effort.

In summary, we answer {\bf RQ2} as follows:

\begin{blockquote}
    In terms of cost versus benefit, we see value in labeling no more than 2.5\% of the data. Using more labels does increase effectiveness but at a much higher cost. Results like this paper can guide managers and developers to make informed decisions about the rewards of performing more work. 
\end{blockquote}

\subsection{\textbf{RQ3: Does the data view in co-training matter?}}
\label{sec:rq5}

\vspace{-0.3mm}
\begin{wrapfigure}{r}{2.5in}
\centering
    \includegraphics[width=\linewidth]{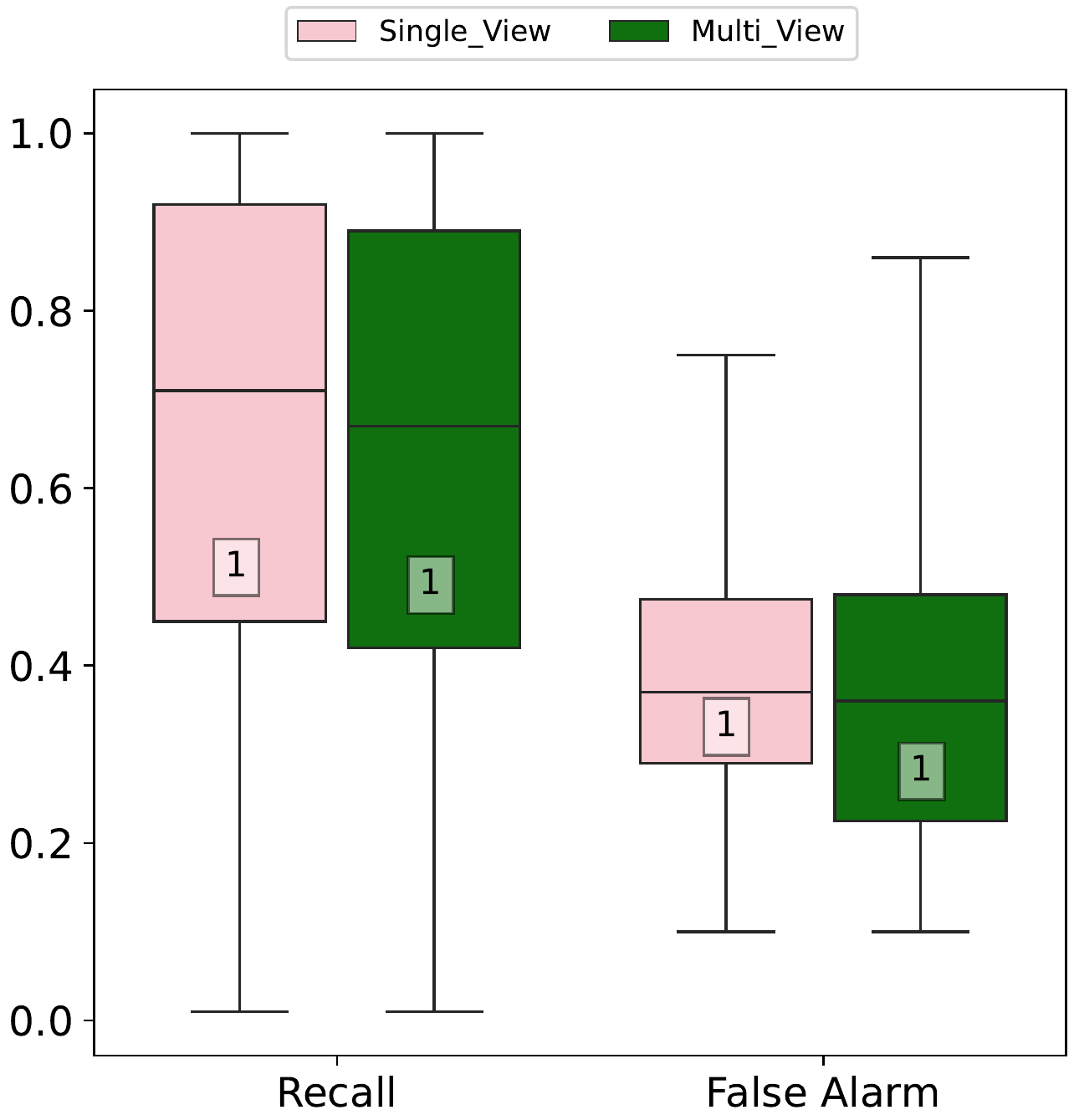}
    \caption{Comparison of recall and false alarm between single\_view and multi\_view co\_training models. The numbers on the box-plots show the statistical rank between single\_view and multi\_view co\_training models.} 
    \label{fig:Single_vs_Multi}
\end{wrapfigure}

Section~\ref{sec:semi-supervised} mentioned that Co-training is a wrapper-based method where two learners are trained on labeled and possibly pseudo-labeled data. While exploring Co-training, we adapt both single-view (same features between models) and multi-view (different features between models) settings. In {\bf RQ3}, we ask if the data views in co-training add significant value to semi-supervised learning.

Figure~\ref{fig:Single_vs_Multi} shows the ranks found in a comparison of different co-training-based models when the training was performed in the multi-view vs. single-view setting. The ``1'' in those plots indicates that the results from both sets of treatments were ranked the same; i.e., we can report that both single-view and multi-view models achieve the same statistical rank, indicating that the performance difference between the two types is not statistically significant.

In summary, we answer  {\bf RQ3} as follows:

\begin{blockquote}
    The choice of different views does not affect the model's performance, and the extra time required to select the different features for the base learners is not justifiable. The extra time required to select the different features for the base learners is not justifiable.
\end{blockquote}

\subsection{RQ4: Does mutual-teaching improved SSL performance compared to self-teaching?}
\label{sec:rq6}

\begin{figure}[!h]
\centering
    \includegraphics[width=.5\linewidth]{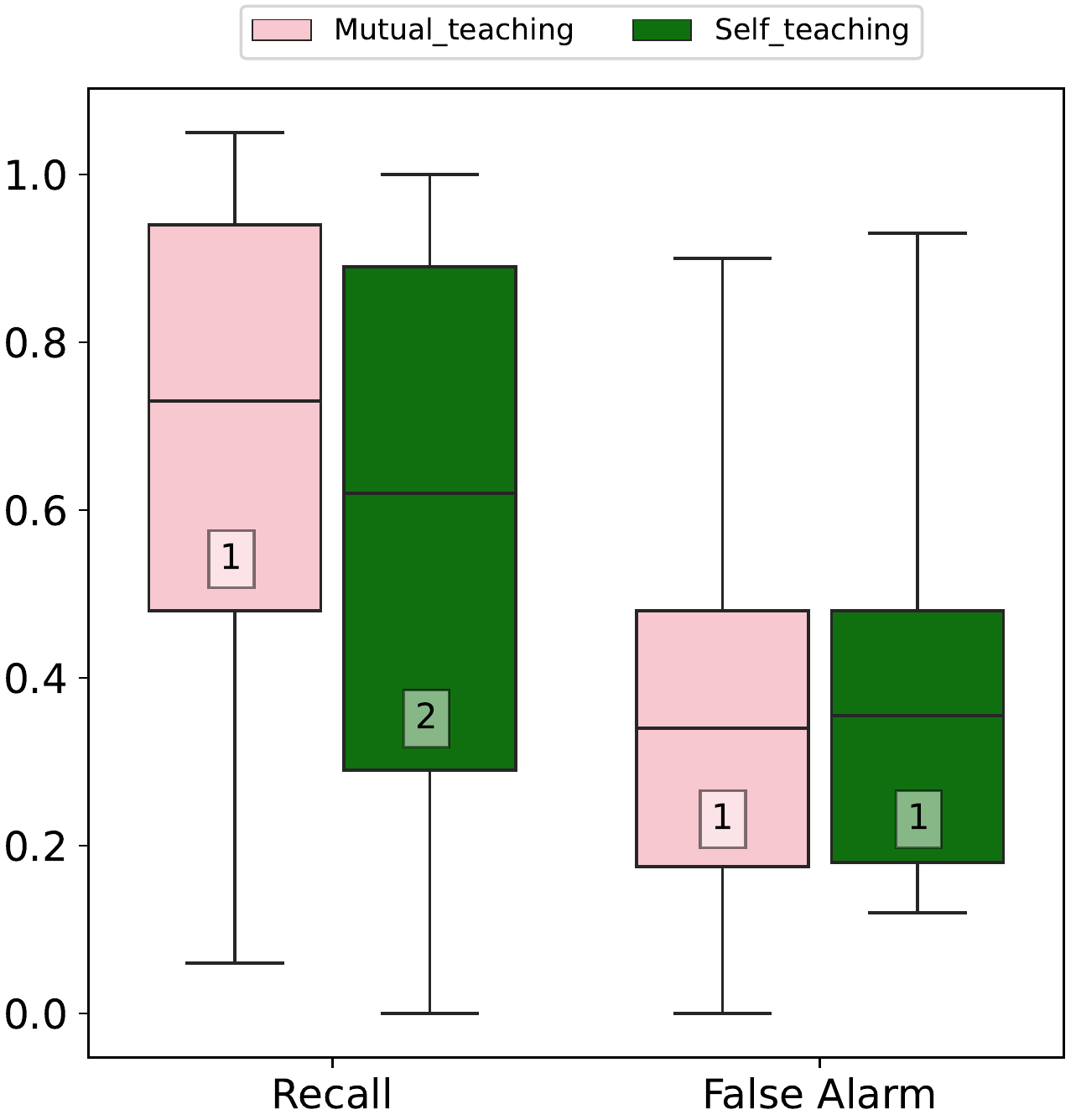}
    \caption{Comparing results between self\_teaching and mutual\_teaching based models.   The number on the box plot shows Scott-Knot rankings.  } 
    \label{fig:self_vs_mutual}
\end{figure}

Figure~\ref{fig:self_vs_mutual} shows results where a learner finds labels from its own previous iteration (self-teaching) or from some other learner (mutual-teaching). While these approaches are statistically similar (in terms of false alarms), we note that  the recall results are statistically better when the learner finds labels from some other learner (mutual-teaching).  Hence, in answer to {\bf RQ4}, we say:

\begin{blockquote}
In semi-supervised learning, it can be more helpful to listen to someone else's opinion than your own.
\end{blockquote}

\section{Discussion}
\label{sec:Discussion}

In this section, we discuss the finding and their implication for future research and verify what makes this research unique.

In our experiments with different SSL methods, we see that in Figure~\ref{fig:supervsied_vs_semisupervised}, semi-supervised methods (that use only some data) can perform better than supervised methods that use all the data. We also see that not all semi-supervised methods perform statistically similarly in terms of recall (see Figure~\ref{fig:Recall}) and false alarm (see Figure~\ref{fig:Pf}). Looking more closely, we see that in {\em recall}, there are no supervised methods in the top rank; in {\em precision}, nearly everything is statistically the same, so those results are uninformative; in {\em false alarm}, there is one supervised method (random forests, or RF) but many, many more unsupervised methods, and finally in {\em G-Score}, it is true that half the supervised methods appear in the top rank. However, two-thirds of the semi-supervised methods are ranked ``top''. Also, a case could be made that G-Score is not so informative since most of the methods tie at the top rank. This indicates that a semi-supervised model, trained with only a small portion of data, is adequate for predicting defects. Following these results, future researchers can utilize appropriate semi-supervised learning methods based on their goal, which will help them create better predictors with only a fraction of the data. 

Our result also shows that the semi-supervised models perform better with increased data availability. Between 2.5 and 100, there is a ``sweet spot'' of 20\% labels, which outperforms 2.5\% SSL (and hence, transitively, 100\% fully-supervised learning). However, with only 2.5\% data available, the semi-supervised models perform reasonably well (achieving the same statistical rank as supervised models, as can be seen in Figure~\ref{fig:supervsied_vs_semisupervised}), significantly reducing potential labeling requirements. Table~\ref{tbl:cost} shows that to run the supervised methods, we would need approximately \$150k for labeling all files while following the semi-supervised learning process with 2.5\%, we would need significantly (40 times) less effort to label, which would require only \$3.8k. Looking at the last row of Table~\ref{tbl:increased_data_size}, we see that the size of the improvement from 2.5\% to 20\% is about ten points improvement in recall and 11 points in false alarm. Between 2.5\% to 20\%, there is a 10\% labeling policy offering some improvements: about seven points improvement in recall and six points in false alarm, while being eight and four times more expensive. It is up to the reader to decide if a four to eight-fold increase in cost is worth  (e.g.)  seven to ten points more recall. However they decide, our case that SSL is useful for SE will persist. This result also shows that using this semi-supervised policy of only 2.5\% labeled data, we can leverage poorly maintained projects by relabeling only a fraction of the data to build a better model.  

Our result also shows that the models' training strategy does affect its performance. When we grouped the models based on training strategy, self-teaching (where the most confident pseudo-labeled examples come from the same learner's previous iteration during the pseudo-labeling phase) and mutual-teaching strategy (where the most confident pseudo-labeled examples come from the other classifier), while these approaches have statistically similar false alarm, the mutual-teaching strategy performs statistically better in-term of recall. We see that mutual-teaching takes advantage of the interplay between two classifiers, where they learn from each other's most confident pseudo-labeled examples. This interplay improves the overall recall performance, allowing the models to recognize defective software modules better. Because mutual-teaching uses pseudo-labels from different classifiers, it leads to a more collaborative and diverse learning process, creating greater robustness to noisy or ambiguous data instances than self-teaching. 

As to how to apply co-training, we know that Co-training models can be built using various strategies, such as single-view (same features between models) and multi-view (different features between models). The key difference between the two models is that in the case of single-view models, both base learner sees the data with the same feature sets. During the pseudo-labeling phase for both base learners, the prediction labels and confidence scores are based on the same features. While for multi-view co-training models, the two base learners are trained on different features. Thus, during the pseudo-labeling phase, when generating the prediction labels and confidence score, the models see completely different features (which should reduce over-fitting), as shown in literature in other domains. However, our result shows that contrary to that belief, seeing the same or different data view in SE defect prediction does not affect the model performance. Selecting different features requires extra time, and spending the extra time to select the different features for the base learners is not justifiable for the software engineering defect prediction domain.

The paper investigates the effectiveness of semi-supervised learning (SSL) for software defect prediction. Through experiments, it is demonstrated that SSL methods outperform fully supervised methods, achieving improved performance with reduced labeling efforts. The study highlights the importance of the training strategy, with the mutual-teaching approach yielding better recall results than self-teaching. Interestingly, using the same features for base learners in co-training models is sufficient for software defect prediction, making the extra effort of selecting different features unnecessary. Overall, the research emphasizes the cost-effectiveness and utility of SSL for defect prediction in software engineering; specifically, researchers can use single-view Co-training models, which have been trained with a mutual-teaching strategy to achieve good performance.

\section{THREATS TO VALIDITY}
\label{sec:threats}

As with any large-scale empirical study, biases can affect the final results. Therefore, any conclusions made from this work must be considered with the following issues in mind:

(a) \textit{Evaluation Bias}: 
In all research questions, we have shown the performance of numerous models built with the process and metrics. We compared them using statistical tests to conclude which is a better and more generalizable predictor for defects. While those results are accurate, that conclusion is scoped by the evaluation metrics we used to write this paper. It is possible that using other measurements, there may be a difference in these different kinds of projects. This is a matter that needs to be explored in future research.  

(b) \textit{Construct Validity}: At various places in this report, we made engineering decisions about the choice of machine learning models,  selecting metric vectors. While those decisions were made using advice from the literature, we acknowledge that other constructs might lead to different conclusions. 

(c) \textit{External Validity}: We have collected data from 714 GitHub Java projects for this study. The process metrics were collected using our code on top of the Commit\_Guru repository. There is a possibility that the calculation of metrics or labeling of defective vs. non-defective using other tools or methods may result in different outcomes. That said, these tools have detailed documentation about the metrics calculations. We have shared our scripts and processes to convert the metrics to a usable format and have described the approach to label defects.  

(d) \textit{Sampling Bias}: We can report three sampling biases in this paper: our choice of data, our choice of SSL methods, and the domain explore:

\begin{itemize}
    \item {\em choice of data:} Our conclusions are based on the 714 projects collected from GitHub. It is possible that different initial projects would have led to different conclusions. That said, this sample is substantial, so we have some confidence that this sample represents an interesting range of projects.  
    \item {\em Problem domain:} Another kind of sampling bias is the problem domain explored here. As stated in the introduction, all our experiments come from the arena of defect prediction~\cite{ghotra2015revisiting, zimm09, okutan2014software, Zhang16aa, nam2018heterogeneous, Tantithamthavorn18, zimmermann09} since that is a very active area of research, and one with much practitioner interest~\cite{tosun2010ai, kim2015remi, wan2018perceptions}. It is an open issue if the conclusions reached in this paper apply to other domains. Based on the results shown here, we strongly urge the research community to repeat the analysis of this paper on those other domains. 
    \item {\em Scope of SSL study:} Since there are so many SSL methods, we had to limit the scope of this investigation (for pragmatic reasons). Hence, we have yet to explore and use all of these semi-supervised methods, as reported by Van et al.~\cite{van2020survey}. For example, we have excluded expensive methods very computationally (since that would complicate a large-scale analysis such as this one). In future work, it would be insightful to explore those excluded methods.
\end{itemize}

\section{Conclusion}
\label{sec:conclusion}
Most techniques for finding defective modules using a defect prediction model are based on supervised classification algorithms, assuming sufficient labeled data is available for training the models. The drawback of supervised classification algorithms is the necessity of having a substantial number of software modules labeled as faulty or fault-free before the model can be created. Labeling is labor-intensive; an alternate approach is to use semi-supervised classification algorithms. However, the application of semi-supervised classification algorithms in defect prediction is far from adequate compared to the number of supervised classification algorithms. Literature review shows a plethora of semi-supervised classification algorithms available in the machine learning domain, and only a few of them have been studied in the defect prediction domain. 

In this paper, we contribute to addressing this gap by conducting an extensive analysis of 55 semi-supervised classification algorithms across eight different groups. This represents one of the largest demonstrations of semi-supervised classification algorithms in the defect prediction domain. Additionally, we compare these algorithms using a dataset of 714 open-source GitHub projects, which is also one of the largest selections of projects used for demonstrating semi-supervised classification algorithms in the defect prediction domain.

The findings of this study reveal that the choice of the learner does matter based on the researcher's goals, providing valuable insights for both researchers and industry practitioners. Moreover, our results demonstrate that employing semi-supervised models can significantly reduce the effort required for labeling as well as be useful to poorly maintained projects while still achieving satisfactory performance. Even with as little as 2.5\% of labeled data, the semi-supervised models perform remarkably well compared to supervised models.

Furthermore, this paper explains the main characteristics and advantages of each semi-supervised classification method, enabling informed decisions based on the specific goals and problem domains. By exploring various variants of semi-supervised algorithms rarely investigated in the defect prediction domain, we expand the understanding of their applicability and potential.

The study also highlights the impact of semi-supervised co-training on model performance and emphasizes the significance of the training and evaluation strategy chosen. The comprehensive analysis presented in this paper provides a valuable resource for future researchers, allowing for the reproducibility and utilization of these models in their studies within the defect prediction domain.

While our findings primarily pertain to the domain of defect prediction, future work can explore the application of this analysis to other domains within software engineering, offering new insights and directions for further research in those areas.

\section{Acknowledgements}
This material is based upon work supported by the National Science Foundation under grant \#1908762. 
Any opinions, findings, conclusions, or recommendations expressed in this material are those of the author(s) and do not necessarily reflect the views of the National Science Foundation.

\section{Declaration}
We wish to confirm that there are no known conflicts of interest associated with this publication, and
there has been no significant financial support for this work that could have influenced its outcome.

\section{Data availability statements}
The datasets used in this study are available at \url{https://github.com/ai-se/Semi-Supervised}


\bibliographystyle{plain}
\bibliography{main}

\end{document}